\documentclass[12pt]{article}
\usepackage{epsfig,amssymb,psfrag}

\newcommand{\beq}{\begin{equation}}
\newcommand{\eeq}{\end{equation}}
\newcommand{\be}{\begin{equation}}
\newcommand{\ee}{\end{equation}}

\def \e  {\mathop{\rm e}\nolimits}
\def \be  {\begin{equation}}
\def \ee  {\end{equation}}
\def \ba  {\begin{eqnarray}}
\def \ea  {\end{eqnarray}}

\def\be{\begin{equation}}
\def\ee{\end{equation}}
\def\bea{\begin{eqnarray}}
\def\eea{\end{eqnarray}}
\newcommand \ci [1] {\cite{#1}}
\newcommand \bi [1] {\bibitem{#1}}
\def \lab #1 {\label{#1}}
\newcommand\re[1]{(\ref{#1})}
\def \qqquad {\qquad\quad}

\newcommand\lr[1]{{\left({#1}\right)}}

\def \Tr {\mbox{Tr\,}}

\newcommand \vev [1] {\langle{#1}\rangle}

\newcommand \ket [1] {|{#1}\rangle}
\newcommand \bra [1] {\langle {#1}|}
\def \CO {{\cal O}}
\font\cmss=cmss10 \font\cmsss=cmss10 at 7pt
\def\inbar{\,\vrule height1.5ex width.4pt depth0pt}
\def\IC{\relax\hbox{$\inbar\kern-.3em{\rm C}$}}
\def\IZ{\relax\ifmmode\mathchoice
{\hbox{\cmss Z\kern-.4em Z}}{\hbox{\cmss Z\kern-.4em Z}} {\lower.9pt\hbox{\cmsss
Z\kern-.4em Z}} {\lower1.2pt\hbox{\cmsss Z\kern-.4em Z}}\else{\cmss Z\kern-.4em
Z}\fi}
\def\IR{{\hbox{{\rm I}\kern-.2em\hbox{\rm R}}}}
\def\IP{{\hbox{{\rm I}\kern-.2em\hbox{\rm P}}}}
\newcommand{\as}{\ifmmode\alpha_{\rm s}\else{$\alpha_{\rm s}$}\fi}
\renewcommand{\Re}{\mathop{\rm Re}\nolimits}

\newcommand \Mybf[1] {\mbox{\boldmath$ {#1} $}}
\newcommand \mybf[1] {\mbox{\boldmath$ {\scriptstyle #1} $}}

\begin{document}

\begin{flushright}
IHES/P/02/21  \\
ITEP-TH-12/02\\
LPTHE-02-19  \\
LPT-Orsay-02-15\\
OUTP-02-06 \\
hep-th/0204183
\end{flushright}

\setcounter{footnote}{0}

\setcounter{equation}{0}
\centerline{
\bf High Energy QCD:
Stringy Picture from  Hidden Integrability}
\vspace{0.3cm}
\centerline{ A. Gorsky$^{\,a,b}$, I.I. Kogan$^{\,a,c,d,e}$ and
G. Korchemsky$^{\,e}$ }

\begin{center}
$^a$ {\em Institute of Theoretical and Experimental Physics, \\
B.Cheremushkinskaya 25, Moscow,  117259, Russia }\\
$^b$ {\em LPTHE, Universit\'e  Paris VI, \\
4 Place Jussieu, Paris, France}\\
$^c$
{\em  Theoretical Physics, Department of Physics, Oxford University\\
1 Keble Road, Oxford, OX1 3NP, UK } \\
$^d$
{\em IHES, 35 route de Chartres \\
91440, Bures-sur-Yvette,  France }\\
$^e$ {\em Laboratoire de Physique Th\'eorique \\ Universit\'e  Paris XI, 91405
Orsay Cedex, France}
\end{center}

\begin{abstract}
We discuss the stringy properties of high-energy QCD using its hidden
integrability in the Regge limit and on the light-cone. It is shown that
multi-colour QCD in the Regge limit belongs to the same universality class as
superconformal $\cal{N}$=2 SUSY YM with $N_f=2N_c$ at the strong coupling
orbifold point. The analogy with integrable structure governing the low energy
sector of $\cal{N}$=2 SUSY gauge theories is used to develop the brane picture
for the Regge limit. In this picture the scattering process is described by a
single M2 brane wrapped around the spectral curve of the integrable spin chain
and unifying hadrons and reggeized gluons involved in the process.
New quasiclassical quantization conditions for the complex higher integrals of
motion are suggested which are consistent with the $S-$duality of the
multi-reggeon spectrum. The derivation of the anomalous dimensions of the lowest
twist operators is formulated in terms of the Riemann surfaces.
\end{abstract}

\section{Introduction}

During the recent years it was recognized that there is the hidden integrability
behind the evolution equations in asymptotically free Yang-Mills
theories~\cite{lipatov,fk}. It was found that high-energy asymptotics of
scattering amplitudes in QCD in the Regge limit are described by evolution
equations which are ultimately related to the $SL(2,\mathbb{C})$ XXX Heisenberg
integrable spin chains. More recently, a similar connection was observed for
another physically interesting limit, namely the QCD dynamics on the light-cone.
There, the logarithmic evolution of the composite light-cone 
 operators is connected with the
real $SL(2,\mathbb{R})$ XXX Heisenberg spin chain and the spectrum of anomalous
dimensions of those operators was identified with the spectrum of this magnet
\cite{BDM,bm,AB,bk}. For the time being it seems that the phenomena is quite
general and the effective QCD dynamics in different kinematical limits is
integrable indeed.

On the other hand, there is another class of (supersymmetric) Yang-Mills theories
in which integrability emerges. In the $\cal{N}$=2 SUSY YM theory which can be solved
in the low-energy limit exactly \cite{sw1} the relevant integrable system was
identified as the complexified periodic Toda chain. The low-energy effective
action and the BPS spectrum in the theory can be described in terms of the
classical Toda system \cite{gkmmm}. When one adds the fundamental matter with
$N_f<2N_c$, which does not change the asymptotically free nature of the theory,
the universality class of integrable model is changing. Instead of the classical
Toda system  one gets the classical XXX spin chain~\cite{gmmm}. In the case of
the superconformal $N_f=2N_c$ theory, the solution leads to the periodic XXX
chain in the magnetic field~\cite{ggm}. Because these magnets are classical, the
spins are not quantized and their values are determined by the matter masses. In
the special case, when the matter is massless, one arrives at the classical XXX
magnet of spin zero.

At the first glance there is nothing in common neither between high-energy
(Regge) limit of QCD and low-energy limit of SUSY YM, nor between the
corresponding integrable models. In this paper we shall demonstrate the opposite
and show that there is a deep relation between the Regge limit of QCD and some
particular limit of superconformal $\cal{N}$=2 SUSY YM at $N_f=2N_c$. This
relation is based on the fact that in the both cases we are dealing with the same
class of integrable models -- the $SL(2,\mathbb{C})$ homogenous spin zero
Heisenberg magnets. However, there is the key difference between two theories.
The integrable model in the SUSY YM is a classical Heisenberg magnet, whereas the
Regge limit of QCD is described by a quantum Heisenberg magnet. To bring two
pictures together, it is natural to develop quasiclassical description in QCD
case and quantize the classical picture in SUSY YM. The first issue was
considered in \cite{kor,kk,K-dual} while some arguments concerning the
quantization of the integrable system related to   SUSY YM theories can be found
in \cite{gor}.

The central role in our approach will be played by Riemann surfaces which appear
naturally as spectral curves  in a description of
classical spin chains. To some extent the Riemann surface fixes the universality
class of the classical integrable model and it defines general solution to the
classical equations of motion. This solution is a finite-gap soliton expressed in
terms of $~\theta-$functions on the Riemann surface. This surface has a natural
complex structure whose moduli are parametrized by the integrals of motion. The
classical dynamics is linearized on the Jacobian of this Riemann surface. In the
case of the SUSY YM these Riemann surfaces encode the expression for the low
energy effective actions \cite{gmmm,ggm}. The moduli of the complex structure are
fixed by the vacuum expectation values of the scalar in $\cal{N}$=2 theory. One gets
Riemann surfaces in the quasiclassical limit of the Schr\"odinger equation for
the quantum spin chain related to QCD in the Regge limit \cite{kk,K-dual}. 
The main goal of this
paper is to use these Riemann surfaces to develop the stringy representation for
the effective dynamics of QCD in the Regge limit exploring the known stringy
realization of low-energy regime of  SUSY YM.

Our starting point is the observation that the Seiberg-Witten curve for the
superconformal  $\cal{N}$=2 SUSY YM theory with gauge group $SU(N_c)$ and the
number of flavours $N_f=2N_c$ coincides with the spectral curve for Heisenberg
spin magnet describing multi-colour QCD in the Regge limit. On the SUSY YM side,
we still have several unidentified parameters. The first one is the rank of the
gauge group $SU(N_c)$. On the QCD side, in the multi-colour limit, we also have
one additional parameter, namely the number of reggeized gluons $N=2,3,...$
exchanged in the $t-$channel~\cite{Bar,KP,Cheng}. It will be shown later that the
correspondence between these two theories implies that the number of reggeized
gluons $N=N_c$. However the superconformal theory has additional freedom due to
the arbitrary value of the bare coupling constant $\tau_{0}=\frac{i}{g^2} +2\pi
\theta$ and the Seiberg-Witten curves are bundled over the complex $\tau_{0}$
plane. On the complex $\tau_{0}-$plane there are singular points at which the
discriminant of the curve vanishes, i.e. the spectral curve becomes degenerate.
It turns out that the Regge limit of the multi-colour QCD is related to one of
these singular points, the so-called strong coupling orbifold point
\cite{argyres,minahan} which is
$S-$dual to another singular point corresponding to the weak coupling $\e^{2\pi
i\tau_0}\rightarrow 0$ regime. Let us emphasize that in the superconformal theory
the effective coupling constant does not coincide with the bare one, because
there are finite perturbative one-loop and nonperturbative instanton corrections
to the low-energy effective action. At the strong coupling regime under
consideration one has to take into account contributions from all numbers of
instantons.

After identification of the universality class one can investigate relevant
$S-$duality properties of both theories, which can be read off from the spectral
curve. In the SUSY case, the $S-$duality is the famous electric-magnetic duality.
The Seiberg-Witten solution follows from the compatibility of the duality with
the renormalization group (RG) flow of the effective action on the moduli space
of the theory. In the QCD case the situation is more subtle since contrary to 
the SUSY case we have to deal with the quantum dynamics of the same integrable 
system and one has to
formulate the precise meaning of the duality at the quantum level. It turns out
that the duality manifests itself in the spectrum of the integrals of
motion~\cite{K-dual}. We formulate a new WKB quantization conditions obeying the
duality property and demonstrate that their solutions are in a good agreement
with the numerical calculations of the spectrum of the three reggeon
system~\cite{kornew}. This strongly support the conjecture that the notion of
duality survives at the quantum level. In result we predict that the quantum
spectrum of the $N-$reggeon system in the multi-colour QCD is parametrized by two
$(N-2)-$dimensional vectors $\vec{n}$ and $\vec{m}$, which have the meaning of
``electric'' and ``magnetic'' quantum numbers.

We reformulate the above mentioned WKB quantization conditions in stringy terms
using the brane realization of the BPS spectrum in superconformal $\cal{N}$=2
SUSY YM theory~\cite{vafa,sen}. Formally it is a simple task to identify the
corresponding geometrical objects however the proper meaning of the ``magnetic''
quantum number in the Regge limit is a subtle point. The duality combined with RG
behaviour give rise to brane realization of the ground state in the SUSY YM
theories. There are two apparently different IIA/M and IIB/F pictures. In the
IIA/M picture the ground state is described by the M5 brane with the world-volume
$R^4 \times \Sigma$ where $\Sigma$ is the spectral curve of the corresponding
integrable system \cite{vafa,witten}. The stable BPS states correspond to the M2
branes ending on this M5 brane. In the IIB/F picture \cite{sen} one has to
consider a D3 brain in the orientifold background and BPS states are described by
the dyonic strings stretched between the D3 brane and orientifold.

In this paper we shall start to develop a similar brane picture for the Regge
limit of the multi-colour QCD. We shall argue that the dynamics of the
multi-reggeon compound states in multi-colour QCD is described in this picture by
a single M2 brane wrapped around the spectral curve of the Heisenberg magnet. The
fact that we are dealing with the {\it quantum\/} magnet implies that the moduli
of the complex structure of the spectral curve can take only quantized values. We
shall present a raw of arguments supporting this picture. Let us note that in the
Regge limit there is a natural splitting of the four-momenta of reggeized gluons
into two transverse and two longitudinal components, which is unambiguously
determined by the kinematics of the scattering process. One can make a Fourier
transform with respect to the transverse momenta and define a mixed
representation: two-dimensional longitudinal momentum space and two-dimensional
impact parameter space for which it is natural to use complex coordinates $z$ and
$\bar z=z^*$. The Riemann surface which the brane is wrapped around lives in this
``mixed'' coordinate-momenta space. Different projections of the wrapped brane
represent hadrons and reggeons. The genus of this surface is fixed by the number
of reggeons. Let us note that multi-reggeon states appear from the summation of
only planar diagrams in QCD and from the point of view of topological expansion
they all correspond to the cylinder-like diagram in a colour space. It
contribution to the scattering amplitude is given by the sum of $N-$reggeon
exchanges in  the $t-$channel with $N=2,3...$. It is amusing that in result we
get a picture where we sum over Riemann surfaces of arbitrary genus $g=N-2$.

Finally, we make some comments on a dual (super)gravity realization of the Regge limit of
multi-colour QCD. Our approach is different from the recent
studies~\cite{polchinski,janik1,giddings} of high-energy scattering in QCD based on the
$AdS_5$ geometry within the AdS/CFT correspondence \cite{maldacena}.
We speculate that a reggeized gluon in QCD can be identified as a singleton in
the $AdS_3$ space.


The paper is organized as follows. In Section 2 we remind how the integrable spin
chains appear in the Regge limit of QCD. In Section 3 we review the main features
of the finite-gap solutions to the classical equation of the motion in the
periodic spin chains which are relevant for our consideration. We also discuss
their quantization using the method of Separated Variables~\cite{Sklyanin}. In
Section 4 we identify the universality class of the Regge limit of multi-colour
QCD by comparing the Riemann surfaces in QCD and $\cal{N}$=2 SUSY YM. Based on
this identification, we make a conjecture about brane realization of QCD in the
Regge limit. In Section 5 we investigate the duality properties of the
multi-reggeon states and provide WKB like description of their spectrum in terms
of Riemann surfaces. In Section 6 we propose the stringy/brane picture for the
description of the anomalous dimensions of the lowest twist operators in QCD. In
Section 7 several comments concerning the supergravity dual description of the
Regge limit are presented. The obtained results are discussed in the Conclusion.

\section{Regge asymptotics in QCD and Heisenberg magnets}

It is well known from the old days of the Regge theory that the scattering
amplitudes of hadrons grow at high-energy. Useful framework for understanding
this phenomenon in QCD comes from the studies of the asymptotic behaviour of the
scattering amplitude of two hadronic states consisting of a pair of heavy
quarks~\cite{Mueller}, the so-called onia system. The advantage of such system is
that it captures all essential features of the Regge phenomenon in QCD and, at
the same time, the scattering amplitudes can be calculated perturbatively
order-by-order in the QCD coupling expansion. Denoting the invariant mass of the
scattering onia by $Q^2$ and their center-of-mass energy by $s$, we introduce the
dimensionless Bjorken scaling variable $x=Q^2/s$.
 The Regge limit corresponds to
the asymptotics at large $s=Q^2/x$ with fixed $Q^2$, or equivalently to small
$x$.

One expects that the calculation of the high-energy asymptotics of the
onium-onium cross-section should eventually match the Regge model prediction. The
Regge model interprets the growth of the scattering amplitudes at high energy
$s$, or equivalently at small $x$, by introducing the notion of the Pomeron as
the Regge pole with the quantum numbers of the QCD vacuum. Its contribution of
the onium-onium cross-section at small $x$ takes the form
\be
\sigma_{\rm tot}(x,Q^2) =  x^{-(\alpha_{ \IP}-1)}\,
\beta_{A}^{ \IP}(Q^2)\, \beta_{B}^{ \IP}(Q^2)\,,
\lab{R}
\ee
with $\alpha_{ \IP}$ the Pomeron intercept and $\beta_{A,B}^{
\IP}(Q^2)$ the residue factors corresponding to the onia $A$ and
$B$, respectively. Despite the fact that the Regge model provides a successful
phenomenological description of the experimental data in terms of ``hard''
(perturbative) and ``soft'' (nonperturbative) Pomerons its status within QCD
remains unclear.

The first attempts to understand the ``hard'' Pomeron within perturbative QCD
have been undertaken more than 20 years ago and they have led to the discovery of
the BFKL Pomeron~\ci{BFKL}. In the BFKL approach, two onia scatter each other by
exchanging soft gluons in the $t-$channel. Calculating the corresponding Feynman
diagrams in powers of $\as$ one obtains the following general form of the
perturbative expansion of the cross-section at small $x$ and fixed
$Q^2$,~\ci{Bar,KP,Cheng}
\be
\sigma_{\rm tot}(x,Q^2)=\sigma_{\rm Born}\sum_{m=0}^\infty\sum_{n=0}^{m}(\as N_c)^{m-n}
(\as N_c \ln x)^{n} f_{m,n}(Q^2)
\,,
\lab{1}
\ee
with $\sigma_{\rm Born}$ being  the cross-section at the Born level. Here, the
double sum involves two parameters of the expansion  -- the QCD coupling
constant, $\as\ll 1$, and the energy dependent parameter $\as\ln 1/x$ that
becomes large at small $x$. The coefficient functions $f_{m,n}(Q^2)$ depend on
the invariant mass of the onia, $Q^2$, as well as on the number of colours,
$1/N_c^2$. They can be calculated perturbatively for large invariant mass $Q^2\gg
\Lambda_{\rm QCD}^2$ by making use of the wave functions of the onia states.
The number of different terms in \re{1} rapidly increases to higher orders in
$\as$ and in order to find $\sigma_{\rm tot}(x,Q^2)$ at small $x$ one has to
develop a meaningful approximation which, firstly, correctly describes the
small$-x$ asymptotics of the infinite series \re{1} and, secondly, preserves the
unitarity constraint
\be
\sigma_{\rm tot}(x,Q^2) < {\rm const.} \times \ln^2 (1/x)
\lab{Fr}
\ee
as $x\to 0$. To satisfy the first condition, one can neglect in \re{1} the terms
containing $f_{m,n}$ with $n<m$ as suppressed by powers of $\as$ with respect to
the leading term $f_{m,m}$. The resulting series defines the onium-onium
cross-section in the leading logarithmic approximation (LLA). It can be resummed
to all orders in $\as$ leading to the BFKL pomeron~\ci{BFKL}
\be
\sigma_{\rm tot}^{\rm LLA}(x,Q^2) = \sigma_{\rm Born}\sum_{m=0}^\infty  (\as N_c\ln x)^m f_{m,m}(Q^2)
\sim \frac{x^{-{\as N_c} 4\ln 2/\pi}}{\sqrt{\as N_c\ln 1/x}}\,.
\lab{BFKL}
\ee
The BFKL Pomeron leads to unrestricted rise of the cross-section and violates the
unitary bound \re{Fr}. In order to preserve the unitarity of the S-matrix of QCD
and fulfill \re{Fr}, one has to take into account an {\it infinite\/} number of
nonleading terms in \re{1}. This means that with the unitarity condition taken
into account the series \re{1} does not have small parameter of expansion like
$\as$. Instead of searching for this parameter one may start with the LLA result
\re{BFKL} and try to identify the nonleading terms in \re{1}, which should be
added to \re{BFKL} in order to restore the unitarity. In this way, one arrives at
the generalized leading logarithmic approximation \ci{Cheng,Bar,KP}. One should
keep in mind, however, that in this approximation the unitarity is preserved only
in the direct channels of the process but not in subchannels corresponding to the
different groups of particles in the final state.

\subsection{Generalized LLA}

In the generalized LLA, one uses the remarkable property of gluon
reggeization~\cite{BFKL} in order to formulate a new diagram technique for
calculation of the onium-onium scattering amplitudes \ci{Cheng,Bar,KP}. Namely,
an infinite set of standard Feynman diagrams involving ``bare'' gluons can be
replaced by a few reggeon diagrams describing propagation of reggeized gluons, or
reggeons, and their interaction with each other. Each reggeon diagram appears as
a result of resummation of an infinite number of Feynman diagrams with ``bare''
gluons. In this way, one defines an effective theory in which the reggeized
gluons start play a role of a new elementary fields. In this theory, the
scattering amplitudes describe the effects of the propagation of the reggeized
gluons in the $t-$channel and their interaction with each other. As usual, each
diagram in the effective theory is equivalent to an infinite sum of the original
QCD Feynman diagrams.

The resulting set of the reggeon diagrams defining the onium-onium cross-section
in the generalized LLA is shown in Figure~1. According to the optical theorem,
the cross-section is given by an imaginary part of their contribution. The
vertical curly lines represent the reggeons propagating in the $t-$channel and
the triple gluon vertices describe the effective BFKL interaction between the
reggeons and the gluons propagating in the $s-$channel. Upper and lower blobs
denote the coupling of the reggeons to two onia states $A$ and $B$.

\begin{figure}[th]
\vspace*{3mm}
\centerline{{\epsfysize4.5cm \epsfbox{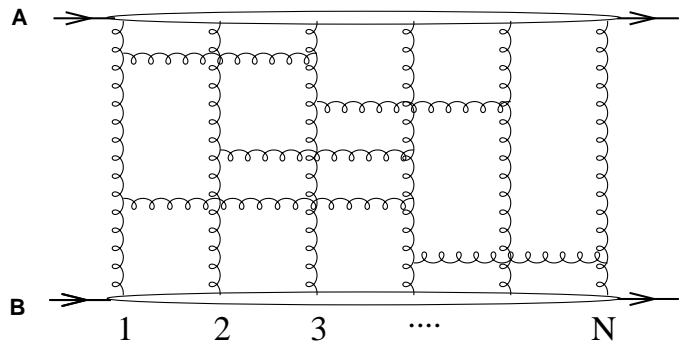}}}
\caption[]{The Feynman diagrams contributing to the scattering amplitude of the
onium-onium scattering in the generalized LLA and describing the $N$ reggeon
exchange in the $t-$channel. }
\label{Fig-N=3}
\end{figure}

One of the peculiar features of the reggeon scattering in the generalized LLA is
that it is elastic and pair--wise. The number of reggeons exchanged in the
$t-$channel is not changed. This allows us to interpret the diagrams shown in
Figure~1 as describing the propagation of the conserved number $N=2,\ 3, ...$ of
pair-wisely interacting reggeons. At $N=2$ the diagram in Figure~1 determines the
leading logarithmic result for the scattering amplitude, Eq.~\re{BFKL}, and it
contributes to the coefficient functions $f_{m,m}(Q^2)$ in \re{BFKL} to all
orders of perturbation theory. The contribution of the diagram with $N=3$ is
suppressed by a power of $\as$ with respect to that for $N=2$ and it determines
the first nonleading coefficient $f_{m,m-1}$ in \re{1}.

The calculation of the $N-$reggeon diagrams can be performed using the
Bartels--Kwiecinski--Praszalowicz approach \ci{Bar,KP}. The diagrams shown in
Figure~1 have a generalized ladder form and for fixed number of reggeons, $N$,
their contribution satisfies Bethe--Salpeter like equations for the scattering of
$N$ particles. The solutions to these equations define the colour-singlet
compound states $\ket{\Psi_N}$ built from $N$ reggeons. These states propagate in
the $t-$channel between two scattered onia states and give rise to the following
(Regge like) high energy asymptotics of the the onium-onium cross-section
\be
\sigma_{\rm tot}(s) \sim \sum _{N=2\,,3,...} (\as N_c)^N \frac{x^{-{\as}N_c \varepsilon_N/\pi}}
{\sqrt{\as N_c\ln 1/x}}\,\beta_A^{(N)}(Q^2)\beta_B^{(N)}(Q^2)
\label{amp}
\ee
with the exponents $\varepsilon_N$ defined below. Here, each term in the sum is
associated with the contribution of the $N-$reggeon compound states. At $N=2$,
the corresponding state defines the BFKL pomeron \re{BFKL} with
$\varepsilon_2=4\ln 2$.

The $N-$reggeon states are defined in QCD as solutions to the Schr\"odinger
equation
\be
{\cal H}_N \Psi(\vec z_1,\vec z_2,...,\vec z_N) = \frac{\as}{\pi}
N_c\varepsilon_N
\Psi(\vec z_1,\vec z_2,...,\vec z_N)
\label{Sch}
\ee
with the effective QCD Hamiltonian ${\cal H}_N$ acting on two-dimensional
transverse coordinates of reggeons, $\vec z_k$  ($k=1,...,N$) and their colour
$SU(N_c)$ charges
\be
{\cal H}_N=-\frac{\as}{2\pi} \sum_{1 \le i < j \le N} H_{ij} \ t_i^a t_j^a \,.
\lab{ham}
\ee
Here, the sum goes over all pairs of reggeons. Each term in this sum is
factorized into the product of two operators acting on the color indices and the
two-dimensional coordinates. The former is given by the direct product of the
gauge group generators in the adjoint representation of the $SU(N_c)$ group,
acting
in the color space of $i-$th and $j-$th reggeons,
$$
t^a_i=\underbrace{ I\otimes ... \otimes t^a}_i \otimes ... \otimes I  \,,
\qqquad  (t^a)_{bc}=-if_{abc}\,,
$$\\[-2mm]
with $f_{abc}$ the structure constants of the $SU(N_c)$. The operator $H_{ij}$
acts only on the transverse coordinates of reggeons, $\vec z_i$ and $\vec z_j$.
The Hamiltonian ${\cal H}_N$ commutes with the total color charge of the system
of $N-$reggeons. Solving \re{Sch}, we are looking for the colour-singlet states,
that is the eigenstates with the total color charge equal to zero,
$\sum_{i=1}^{_N} t^a_i \ket{\Psi_N}=0$. The residue factors in \re{amp} measure
the overlap of $\Psi(\vec z_1,\vec z_2,...,\vec z_N)$ with the wave functions of
two scattered particles, $\beta_{A}^{_{(N)}}(Q^2)=\vev{A|\Psi_N}$ and similar for
$\beta_B^{_{(N)}}$.

To get some insight into the properties of the $N-$reggeon states, it proves
convenient to interpret the Feynman diagrams shown in Figure~1 as describing a
quantum-mechanical evolution of the system of $N$ particles in the $t-$channel
between two onia states $\ket{A}$ and $\ket{B}$
\be
\sigma_{\rm tot}(s)=\sum_{N\ge 2} (\as N_c)^N \vev{A|\e^{\ln (1/x) \cdot{\cal H}_N}|B}\,,
\label{time}
\ee\\[-2mm]
with the rapidity $\ln x=\ln Q^2/s$ serving as Euclidean evolution time. To find
the high energy asymptotics of \re{time}, one has to solve the Schr\"odinger
equation \re{Sch} for arbitrary number of reggeized gluons $N$, expand the
operator $(1/x)^{{\cal H}_N}$ over the complete set of the eigenstates of ${\cal
H}_N$ and, then, resum their contribution for arbitrary $N$. In this way, one
arrives at \re{amp}.

It follows from \re{time} that for fixed number of the reggeons, $N$, and large
evolution time, $\ln 1/x\gg 1$, the dominant contribution to $\sigma_{\rm
tot}(s)$ comes from the eigenstates of ${\cal H}_N$ with the maximal energy. This
allows us to identify the exponent $(-\varepsilon_N)$ in \re{amp} as the energy
of the ground state of the Hamiltonian $(-{\cal H}_N)$. It turns out that the
energy spectrum of the reggeon Hamiltonian is gapless and the contribution of the
eigenstates next to the ground state amounts to the appearance of the additional
factor $\sim(\as\ln 1/x)^{-1/2}$ in the r.h.s.\ of \re{amp}. The contribution of
the $N-$reggeon state to $\sigma_{\rm tot}(s)$, Eq.~\re{amp}, depends crucially
on the sign of the energy -- for $\varepsilon_N>0$ (or $\varepsilon_N<0$) it
increases (or decreases) as $s\to\infty$.

Finding the spectrum of the compound states \re{Sch} for arbitrary number of
reggeized gluons $N$ and eventual resummation of their contribution to the
scattering amplitude \re{amp} is the outstanding theoretical problem in
high-energy QCD. At $N=2$, the solution to \re{Sch} has been found a long time
ago -- the well-known BFKL Pomeron~\cite{BFKL}, $\varepsilon_2=4\ln 2$. At $N=3$
the solution to \re{Sch} -- the Odderon state in QCD \cite{LN}, was formulated
only a few years ago~\cite{janik2}, $\varepsilon_3=-0.2472..$, by making use of
the remarkable properties of integrability of the effective QCD Hamiltonian
\cite{lipatov,fk}. Recently, a significant progress has been achieved in solving
\re{Sch} for arbitrary number of reggeons in the milti-colour limit
\cite{KKM,devega,kornew}.

\subsection{Large $N_c$ limit}

At large $N_c$ one can expand the matrix elements in the r.h.s.\ of \re{time} in
powers of $1/N_c^2$ and obtain a topological expansion of $\sigma_{\rm tot}(s)$
\cite{Ven}. The leading term of this expansion corresponds to the planar cylinder
diagram in which the top and the bottom disks correspond to two onia states and
the walls are formed by $N$ reggeons. In this diagram, in order to preserve the
planarity, each reggeon is allowed to interact only with two neighbouring
reggeons. Going back from the planar diagrams to the Hamiltonian ${\cal H}_N$,
this leads to simplification of the reggeon interaction in the multi-colour
limit. Namely, for $\as N_c={\rm fixed}$ and $N_c\to\infty$, the Hamiltonian
\re{ham} can be written as
\be
{\cal H}_N=\frac{\as N_c}{4\pi} \sum_{1 \le i \le N} H_{i,i+1}+ \CO(N_c^{-2})
\label{Ham-sim}
\ee
with $H_{N,N+1}\equiv H_{N,1}$. Thus, the pair-wise interaction between reggeons
in \re{ham} is reduced in the multi-colour limit to a nearest-neighbour
interaction \re{Ham-sim} and, in addition, the colour operator is replaced by a
$c-$valued factor $t^a_it^a_j \to -N_c/2$. As was already mentioned, the
two-particle reggeon Hamiltonian $H_{i,i+1}$ acts on the two-dimensional
(transverse) reggeon coordinates $\vec z=(x,y)$ and it coincides with the BFKL
kernel~\cite{BFKL}. It becomes convenient to introduce the complex valued
(anti)holomorphic coordinates, $z=x+iy$ and $\bar z=x-iy$, and parameterize the
position of the $k$th reggeon as $\vec z_k=(z_k,\bar z_k)$. One of the remarkable
features of the BFKL kernel is that it is given by the sum of two mutually
commuting operators acting separately on $z-$ and $\bar z-$coordinates of the
reggeons. Making use of this property, one can rewrite the $N-$reggeon
Hamiltonian \re{Ham-sim} as~\cite{BFKL}
\be
{\cal H}_N = \frac{\as N_c}{4\pi}\lr{H_N + \bar H_N} + \CO(N_c^{-2})\,.
\lab{RH}
\ee
Here the hamiltonians $H_N$ and $\bar H_N$ act on the (anti)holomorphic
coordinates and describe the nearest--neighbour interaction between $N$ reggeons
$$
H_N = \sum_{m=1}^N H(z_m,z_{m+1})\,,\qquad
\bar H_N = \sum_{m=1}^N H(\bar z_m,\bar z_{m+1})\,,
$$
with the periodic boundary conditions  $z_{k+1}=z_1$ and $\bar z_{k+1}=\bar z_1$.
The interaction Hamilonian between two reggeons with the coordinates $(z_1,\bar
z_1)$ and $(z_2,\bar z_2)$ in the impact parameter space, is given by the BFKL
kernel
\be
H(z_1,z_2)=-\psi(J_{12})-\psi(1-J_{12})+2\psi(1)\,,
\lab{psi}
\ee
where $\psi(x)=d\ln\Gamma(x)/{dx}$ and the operator $J_{12}$ is defined as a
solution to the equation
\be
J_{12} (J_{12}-1) = -(z_1-z_2)^2\partial_1\partial_2
\lab{J}
\ee
with $\partial_m=\partial/\partial z_m$. The expression for $H(\bar z_1,\bar
z_2)$ is obtained from \re{psi} by substituting $z_k\to\bar z_k$.

Since the Hamiltonian $H_N$ and $\bar H_N$ act along two different directions on
the two-dimensional plane, they commute with each other, $[H_N,
\bar H_N] = 0$. This allows us to reduce the original Schr\"odinger equation
\re{Sch} to the system of two one-dimensional Schrodinger equations for the
Hamiltonians $H_N$ and $\bar H_N$. For fixed number of reggeons, $N$, the
operator $H_N$ describes the system of $N$ particles on a complex $z-$line
interacting with their neighbours through the two-particle hamiltonian \re{psi}.
This quantum-mechanical system can be solved exactly for the system with $N=2$
particles leading to the BFKL Pomeron. It is not obvious, however, whether the
exact solution can be found for an arbitrary number of reggeons. For this to
occur, the system of $N$ reggeons should possess an additional symmetry. It turns
out that such a hidden symmetry indeed exists~\cite{lipatov,fk}. Namely, the
Schr\"odinger equation for the $N-$reggeon system in multi-colour limit contains
the set of the mutually commuting integrals of motion $q_k$ and $\bar q_k$
$(k=2,...,N)$ defined as
\be
q_k=i^k\sum_{1\le j_1 < j_2 < ... < j_k\le N}
z_{j_1j_2}...z_{j_{k-1},j_k}z_{j_k,j_1}\partial_{z_{j_1}}...
\partial_{z_{j_{k-1}}}\partial_{z_{j_k}}
\label{q}
\ee
with $z_{jk}\equiv z_j-z_k$. The charges $\bar q_k$ are given by similar
expressions in the $\bar z-$sector. The eigenstates $\Psi(\vec z_1,\vec
z_2,...,\vec z_N)$ have to diagonalize these operators and the corresponding
eigenvalues $q\equiv\{q_k,\bar q_k=q_k^*\}$, with $k=2,...,N$, form the complete
set of quantum numbers parameterizing the spectrum of the Schr\"odinger equation
\re{Sch}, $\varepsilon_N=\varepsilon_N(q)$. This implies that the Schr\"odinger
equation \re{Sch} for the $N-$reggeon states is completely integrable in the
multi-colour limit and it can be solved by applying the powerful Quantum Inverse
Scattering Method well-known in the theory of integrable models.

The famous example of the exactly solvable many-body quantum mechanical system is
the one-dimensional XXX Heisenberg chain of $N$ interacting spins described by
the hamiltonian
$$
H_N^{{s=1/2}} = - \sum_{m=1}^N \vec S_m \cdot \vec S_{m+1}\,,
$$
where $\vec S_m$ are the $SU(2)$ generators of the spin $s=1/2$ acting in the
$m-$th site and $\vec S_{N+1}\equiv \vec S_1$. It turns out \ci{fk,KKM} that this
simple model admits nontrivial generalizations for an arbitrary complex value of
the spin $s$. In the latter case, the spin operators $\vec S_m$ are the
generators of an infinite-dimensional representation of the $SL(2,\mathbb{C})$
group of the principal series. Remarkably enough, in the special case of the
$SL(2,\mathbb{C})$ spin $s=0$, the Hamiltonian of such completely integrable
noncompact XXX Heisenberg magnet becomes identical to the QCD Hamiltonian
\re{RH}, describing the interaction between $N-$reggeons in the multi-color
limit.

To explain this correspondence let us consider the one-dimensional lattice with
periodic boundary conditions and with the number of sites, $N$, equal to the
number of Reggeons. Each site is parameterized by the two-dimensional vector
$\vec z_m=(z_m,\bar z_m)$ with $z_m$ and $\bar z_m$ being holomorphic and
antiholomorphic coordinates, respectively, and $m=1,..,N$. In addition, one
assign to each site the spins operators $S_k^{\pm,0}$ and $\bar S_k^{\pm,0}$,
which are the six generators of the principal series of the $SL(2,\mathbb{C})$
group. They are realized on the quantum space of the model as the differential
operators
\be
S^0_k=z_k\partial_{z_k} +s\,,\qquad S_k^-=-\partial_{z_k}\,,\qquad
S_k^+=z_k^2\partial_{z_k}+2sz_k\,.
\label{spins}
\ee
The operators $\bar S_k^{\pm,0}$ are given by similar expressions with $z_k$, $s$
replaced by $\bar z_k$, $\bar s=1-s^*$, respectively. The pair of complex
parameters $(s,\bar s)$ specifies the $SL(2,\mathbb{C})$ representation. For the
principal series of the $SL(2,\mathbb{C})$ they are parameterized by integer
$n_s$ and real $\nu_s$ as $s=(1+n_s)/2+i\nu_s$ and $\bar s=1-s^*$. To match the
reggeon Hamiltonian, Eq.~\re{Sch}, the spins have to be equal to $s=0$ and $\bar
s=1$. In this representation, the BFKL kernel defining the interaction between
two reggeons describes the interaction between the nearest spins. The Hamiltonian
of exactly solvable noncompact $SL(2,\mathbb{C})$ Heisenberg magnet of spin
$(s,\bar s)$ is defined as \ci{fk,KKM}
\be
H^{(s)}_N=\sum_{m=1}^N H_{m,m+1}
\,,\qquad H_{m,m+1}=-i\frac{d}{d u} \ln R_{m,m+1}(u,u)\bigg|_{u=0}
\lab{X}
\ee
where the operator $R_{m,m+1}$, the so-called fundamental $R-$matrix for the
$SL(2,\mathbb{C})$ group, satisfies the Yang-Baxter equation and it is given by
\ci{KKM}
\be
R_{12}(u,\bar u)=
\frac{\Gamma(i\bar u)
\Gamma(1+i\bar u)}{
\Gamma(-i u)\Gamma(1-i u)}\,\times\,
\,\frac{\Gamma(1-\bar J_{12}-i\bar u )\,\Gamma(\bar J_{12}-i \bar u)}{
\Gamma(1-J_{12} +i u)\,\Gamma( J_{12} +i u)}\,,
\label{R-homo}
\ee
with $J_{12}$ defined as the sum of two spins
\be
J_{k,k+1}(J_{k,k+1}-1)=(S^{(k)}+S^{(k+1)})^2
\label{J1}
\ee
with $S^{(N+1)}_\alpha=S^{(1)}_\alpha$, and $\bar J_{k,k+1}$ is defined
similarly. The operator $R_{12}(u,\bar u)$ acts on the tensor product $V\otimes
V$ with $V\equiv V^{(s,\bar s)}$ being the representation space of the principal
series of the $SL(2,\mathbb{C})$.

Substituting \re{R-homo} into \re{X}, we verify that the Hamiltonian of
noncompact Heisenberg magnet of the $SL(2,\mathbb{C})$ spin $s=0$ and $\bar s=1$
coincides with the Hamiltonian of Reggeon interaction \re{psi}. This leads to the
following identity between two Hamiltonians
%
\be
{\cal H}_N \equiv \frac{\as N_c}{4\pi}H_N^{(s=0)}+{\cal O}(1/N_c^2)\,.
\lab{=}
\ee
Thus, the problem of finding the spectum of the $N-$reggeon states in
multi-colour QCD is reduced to solving the Schr\"odinger equation for a peculiar
completely integrable model -- the noncompact Heisenberg magnet with spin
operators being the generators of the principal series of the $SL(2,\mathbb{C})$
group of the spin $s=0$. This model has  a number of remarkable properties
\cite{fk,KKM}, which make it completely different from conventional Heisenberg
spin magnets studied so far. Firstly, the $SL(2,\mathbb{C})$ representation of
the principal series does not have the highest weight and, as a consequence, the
Algebraic Bethe Ansatz is not applicable to construction the eigenstates of the
model. Secondly, despite the fact that the Hamiltonian \re{RH} can be split into
the sum of holomorphic and antiholomorphic mutually Hamiltonians acting on the
$z-$ and $\bar z-$coordinates of the particles, the two sectors are not
independent. The interaction between two sectors occurs through the condition for
the wave function to be a single valued function on the $\vec z=(z,\bar
z)-$plane. This requirement plays a crucial role in establishing the quantization
conditions on the integrals of motion and in finding the energy spectrum of the
model \cite{kornew}.

\section{Finite gap solutions 
and their quantization}

To get some insight into the properties of the Schr\"odinger equation \re{Sch},
it becomes convenient to consider a classical analog of the noncompact Heisenberg
spin magnet. As was shown in \cite{kor,kk}, the corresponding classical
Hamiltonian model naturally appears in the WKB solutions of the Schr\"odinger
equation \re{Sch}. {}From point of view of classical dynamics, the system
describes two copies of one-dimensional spin chains ``living'' on the complex
$z-$ and $\bar z-$lines. Each of these spin chains can be analyzed separately
and, for simplicity, we shall concentrate in this Section on the holomorphic
system.

\subsection{Classical dynamics}

Following the Quantum Inverse Scattering Method, the classical homogenous spin
$s=0$ chain of the length $N$ is defined in the holomorphic sector by the
$2\times 2$ Lax matrices
\be\label{LaxXXX}
L_k(x) = x\cdot {\bf 1} + \sum_{a=1}^3 S^a_{k}\cdot\sigma^a
=\lr{\begin{array}{cc}x-iz_k p_k & ip_k \\ -iz_k^2 p_k &x+iz_k p_k
\end{array}}
\ee
where $\sigma^a$ are the Pauli matrices and $x$ is the spectral parameter. Here,
$S^a_i$ stand for classical spins. They are obtained from the spin operators in
the holomorphic sector, Eq.~\re{spins}, by substituting the derivatives with
respect to $z-$coordinates by the corresponding classical momenta
$i\partial_{z_k} \to p_k$
\be\label{q-class}
q_k\bigg|_{\rm classical}=\sum_{1\le j_1 < j_2 < ... < j_k\le N}
z_{j_1j_2}...z_{j_{k-1},j_k}z_{j_k,j_1}p_{j_1}... p_{j_{k-1}}p_{j_k}
\ee
In this way, one can define on the phase space of the model the Poisson brackets
\be
 \{ z_k,p_j\}=\delta_{jk}
\ee
and verify that the Poisson brackets of the dynamical variables $S_a$, $a=1,2,3$
(taking values in the algebra of functions) look like
\be\label{Scomrel}
\{S^a_k,S^b_j\} = -i\epsilon_{abc} S^c_k\,\delta_{jk}\,,
\ee
so that $\{S_a\}$ plays the role of angular momentum (``classical spin'') giving
the name ``spin-chains'' to the whole class of systems. Algebra (\ref{Scomrel})
has an obvious Casimir function $
{\bf S}^2 = \sum_{a=1}^3 S^a_jS^a_j\,,
$
which does not depend on the number of the site, $j=1,...,N$ for the homogenous
spin chain. In similar manner, one can define the classical analogs of the
operators $q_k$ and the Hamiltonian ${\cal H}_N$.

The equations of motion of the classical spin magnet defined in this way look
like
\be
\frac{\partial z_n}{\partial t}
=\{z_n,{\cal H}_N\}=\frac{\partial {\cal H}_N} {\partial p_n}\,,
\qquad
\frac{\partial p_n}{\partial t}
=\{p_n,{\cal H}_N\}=-\frac{\partial {\cal H}_N} {\partial z_n}\,.
\label{EE}
\ee
They possess the set of $N-1$ conserved charges $q_n$
\be
\partial_{t} q_n =\{q_n,{\cal H}_N\}=0\,.
\ee
with $n=2,...,N$ and their solutions define the reggeon trajectories $z_k=z_k(t)$
subject to the periodicity condition $z_{k+N}(t)=z_k(t)$ and $p_{k+N}(t)=p_k(t)$.

To demonstrate a complete integrability of the evolution equations, one observes
that the system \re{EE} is equivalent to the matrix Lax pair relation
\be
\partial_{t} L_k(x) = \{L_k(x),{\cal H}_N\}
=A_{k+1}(x) L_k(x) - L_k(x) A_k(x)\,,
\label{LA}
\ee
where $A_k(x)$ is a $2\times 2$ matrix depending on the momenta and the
coordinates of the particles. Let us introduce the Baker-Akhiezer function
$\Psi_k(x;t)$ as a solution of the following system of matrix relations
\be
\label{T-def}
L_k (x)\, \Psi_k(x,t) = \Psi_{k+1}(x,t)\,,\qquad
\partial_{t} \Psi_k(x,t) = A_k(x)\, \Psi_k(x,t)\,.
\ee
The two-component Baker-Akhiezer function defined in this way depends on the site
number $k$. The Lax operator $L_k (x)$ shifts it into the neighbouring site. One
can introduce the monodromy operator producing the shift of $\Psi_k$ along the
whole chain as
\be
\label{Tmat}
\Psi_{k+N}(x;t)=T_N(x)\Psi_{k}(x;t)\,,\qquad T_N(x)\equiv L_N(x)\ldots L_1(x)
\ee
The periodic boundary conditions on the solutions to \re{EE} can be easily
formulated in terms of the Baker-Akhiezer function  as
\be
\label{pbc}
\Psi_{k+N}(x;t)= w \,\Psi_{k}(x;t)
\ee
where $w$ is the Bloch-Floquet factor. According to \re{Tmat} and \re{pbc}, $w$
is the eigenvalue of the monodromy operator $T_N(x)$, and therefore it satisfies
the characteristic equation
\be
\label{specurv}
\det (T_N(x)-w\cdot {\bf 1})=0
\ee
{}From Eqs.~\re{LA} and \re{Tmat}, the time dependence of the monodromy operator
is given by $\partial_t T_N(x)= A_1(x) T_N(x)-T_N(x) A_1(x)$, and, as a
consequence, its eigenvalues $w$ are conserved quantities generating a complete
set of integrals of motion. Evaluating the l.h.s.\ of \re{specurv}, one obtains
the spectral curve  equation
\be\label{scsl2m}
w^2 - w \cdot\Tr T_N(x) + \det T(x)=0\,.
\ee
Using the definitions \re{T-def} and \re{LaxXXX} one gets
\be
t_N(u)\equiv \Tr T_N(x)= 2x^N + q_2 x^{N-2} + ... + q_{N-1} x + q_N
\label{t_N}
\ee
with $q_k$ being the integrals of motion, Eq.~\re{q-class}, and
\be
\det T(x)=\prod_{k=1}^N \det L_k(x) = x^{2N}\,.
\ee
Introducing the complex function $y(x)=w-x^{2N}/w$ one obtains the equation of
the spectral curve in the form
\be
\Gamma_N: \qquad
y^2=t_N^2(x)-4x^{2N}=(4x^N + q_2 x^{N-2} + ... + q_N)(q_2 x^{N-2} + ... + q_N)\,.
\label{Gamma_N}
\ee
For any complex $x$ in general position the equation \re{scsl2m} has two
solutions for $w$, or equivalently $y(x)$ in \re{Gamma_N}.

Under appropriate boundary conditions on $\Psi_k(x)$ these solutions define two
branches of the Baker-Akhiezer function, $\Psi^\pm_k(x)$. Then, being a
double-valued function on the complex $x-$plane, $\Psi_k(x)$ becomes a
single-valued function on the Riemann surface corresponding to the complex curve
$\Gamma_N$. This surface is constructed by gluing together two copies of the
complex $x-$plane along the cuts $[e_2,e_3]$, $...$, $[e_{2N-2},e_1]$ running
between the branching points $e_j$ of the curve \re{Gamma_N}, defined as simple
roots of the equation
\be
t_N^2(e_j)=4e_j^{2N}\,,
\qquad
j=1,...,2N-2.
\ee
Here  positions on the complex plane depend on the values of the integrals of
motion $q_2$, $q_3$, .$..$, $q_N$. In general, the Riemann surface defined in
this way has a genus $g=N-2$ which depends on the number of reggeons, $N$. For
instance, it is a sphere at $N=2$ and a torus at $N=3$.

Applying the standard methods of the finite-gap theory, it becomes possible to
construct the explicit expression for the Baker-Akhiezer function in terms of the
theta-functions defined on the Riemann surface \re{Gamma_N} and, finally, obtain
the solutions to the classical equations of motion \re{EE}. The explicit
expressions can be found in \cite{kk}. So far we restricted our consideration to
the classical dynamics in the $z-$sector. Obviously, similar relations hold in
the $\bar z-$sector. Since the integrals of motion, $q_k$ and $\bar q_k$, and the
Hamiltonians in two sectors, $H_N$ and $\bar H_N$, are   conjugated to each
other, the solutions of the classical equations of motion in two sectors can be
obtained one from another.

Combining together the classical motion along the $z-$ and $\bar z-$directions, we
find that the resulting expressions describe the finite-gap soliton wave
propagating in the system of $N$ particles located on the two-dimensional plane.
In terms of the spectral curve \re{Gamma_N}, the same classical dynamics
corresponds to the motion of  particles on the Riemann surface \re{Gamma_N}.
The motion on the Riemann surface is not confined to the kinematically allowed
bands corresponding to the $\alpha-$cycles and the classical trajectories wrap
along both the $\alpha-$ and $\beta-$cycles. This is one of the manifestation of
the fact that we are dealing with two-dimensional system. As we see below, it
will play an important r\^ole in our analysis.

\subsection{Quantum case}

Let us now consider the quantum evolution of the system with the Hamiltonian
defined in \re{X}. As was explained in the Section 2.1, the eigenstates of the
noncompact $SL(2,\mathbb{C})$ Heisenberg chain of length $N$ and spin $(s=0,\bar
s=1)$ play the role of the wave functions of the colourless compound states of
$N$ reggeized gluons. In addition, the ground state of the system controls the
high-energy asymptotics of the scattering amplitudes, Eq.~\re{amp}.


Due to complete integrability of the system, the $N-$Reggeon states have to
diagonalize simultaneously the integrals of motion $q_k$ and $\bar q_k$ with
$k=2,...,N$  Their corresponding eigenvalues  provide the total set of the
quantum numbers of the $N-$Reggeon states, $(q,\bar q)$. In particular, the
``lowest'' integrals of motion, $q_2$ and $\bar q_2$, coincide with the quadratic
Casimir operators for the total $SL(2,\mathbb{C})$ spin of the system and their
eigenvalues are related to the conformal $SL(2,\mathbb{C})$ spin of the
$N-$Reggeon state as
\be
q_2=-h(h-1) \,,\qqquad
\bar q_2 =-\bar h (\bar h-1)\,,\qqquad
\bar q_2=q_2^*\, , 
\label{q2}
\ee
where all  possible values of $h$ and $\bar h=1-h^*$ can be parameterized by
integer $n$ and real $\nu$
\be
h=\frac{1+n}2 + i\nu\,,\qquad n=\IZ\,,\quad \nu = \IR\,.
\lab{h}
\ee
As to remaining charges, $q_3$, ..., $q_N$ and $\bar q_k=q_k^*$, their possible
values are also quantized. The explicit form of the corresponding quantization
conditions is more complicated and it was obtained in \ci{kornew}.

According to \re{q}, the integrals of motion are given by the $k-$order
differential operators (with $k=2,...,N$) acting on the holomorphic and
antiholomorphic reggeon coordinates. Therefore, their spectral problems leads to
a complicated system of coupled differential equations on the wave functions of
the $N-$reggeon states $\Psi_{q,\bar q}(\vec z_1,...,\vec z_N)$. The solution to
this system can be found by applying the method of separated variables developed
by Sklyanin \cite{Sklyanin}. 
Namely, there exists the unitary transformation $U_{\vec p,\vec
x}(\vec z)$
\be
U_{\vec p,\vec x_1,...,\vec x_{N-1}}(\vec z_1,...,\vec z_N) =
\vev{\vec z_1,...,\vec z_N|\vec p,\vec x_1,...,\vec x_{N-1}}\,,
\label{SoV-unit}
\ee
that allows us to transform the wave function from the original $\vec
z-$representation defined by the $N$ coordinates of Reggeons on the
two-dimensional plane to the representation of the separated coordinates
$\vec{{x}}=(\vec x_1,...,\vec x_{N-1})$ and $\vec p$ being the total momentum of
the system. The eigenstate of the Hamiltonian in this representation is given by
\ba
\Phi_{\{q,\bar q\}}(\vec x_1,...,\vec x_{N-1})\delta(\vec p-\vec p\,')
&=&\vev{\Psi_{\vec p,\{q,\bar q\}}|\vec p\,',\vec x_1,...,\vec x_{N-1}}
\nonumber
\\
&=&
\int d^2{z} \, U_{\vec p\,',\vec x_1,...,\vec x_{N-1}}(\vec z)\,
\Psi_{\vec p,\{q,\bar q\}}(\vec z)
\,,
\label{Phi-def}
\ea
where $d^2z=d^2z_1 ... d^2 z_N$. The remarkable feature of the unitary
transformation $U_{\vec p,\vec x}(\vec z)$ is that it transforms the wave
function $\Psi_{\vec p,\{q,\bar q\}}(\vec z)$ satisfying the original
Schr\"odinger equation \re{Sch} into the wave function in the separated
coordinates, $\Phi_{\{q,\bar q\}}(\vec x)$, which is factorized into the product
of $Q-$functions depending on a single separated coordinate $\vec x_k$ and
satisfying the Baxter equation (see Eqs.~\re{Bax1} and \re{Bax2} below)
\be
\Phi_{\{q,\bar q\}}(\vec x_1,...,\vec x_{N-1})
= 
Q(x_1,\bar x_1) \ldots Q(x_{N-1},\bar x_{N-1})\,.
\label{prod-Q}
\ee
Here, the notation was introduced for the holomorphic and antiholomorphic
components of the separated coordinates $\vec x_k=(x_k,\bar x_k)$. The explicit
construction of the transformation to the separated coordinates,
Eq.~\re{Phi-def}, can be found in \cite{KKM}. Requiring $U_{\vec p\,',\vec
x_1,...,\vec x_{N-1}}(\vec z)$ to be a single-valued function on the $\vec
z-$plane, one finds that the possible values of the separated coordinates $\vec
x_k$ have the following form
\be
x_k=\nu_k-\frac{in_k}2\,,\qquad \bar x_k=\nu_k+\frac{in_k}2
\label{x-quan}
\ee
with $\nu_k$ real and $n_k$ integer.

The functions $Q(x,\bar x)$ entering \re{prod-Q} have to satisfy the functional
Baxter equations in the $x-$sector
\be
x^N Q(x+i,\bar x)+ x^NQ(x-i,\bar x)=t_N(x) Q(x,\bar x)
\label{Bax1}
\ee
and similar relation holds in the $\bar x-$sector
\be
(\bar x+i)^N Q(x+i,\bar x)+ (\bar x-i)^N Q(x-i,\bar x)=\bar t_N(\bar x) Q(x,\bar
x)\,.
\label{Bax2}
\ee
Here, $t_N(x)$ is a polynomial of degree $N$ in $x$ with the coefficients given
by the eigenvalues of the integrals of motion, Eq.~\re{t_N}. $\bar t_N(\bar x)$
is given by similar expression with $q_k$ replaced by $\bar q_k=q_k^*$. To make
the correspondence with the classical mechanics, one introduces the shift
operator on the space of functions $Q(x,\bar x)$
\be
\e^{\pm p} Q(x,\bar x)= Q(x\pm i,\bar x)\,,
\qquad \e^{\pm \bar p} Q(x,\bar x)= Q(x,\bar x\pm i)\,.
\ee
Then, the Baxter equations can be rewritten as
\be
\left[x^N \lr{ \e^p +\e^{-p}} - t_N(x)\right] Q(x,\bar x)
=\left[\bar x^N \lr{ \e^{\bar p} +\e^{-\bar p}} - \bar t_N(\bar x)\right]
\bar x^N Q(x,\bar x)=0\,.
\label{Bax-eq}
\ee
Based on the Baxter equations, it is now possible to establish the correspondence
between the quantum and the classical systems. Let us forget for a moment that
$p$ and $x$ are operators in \re{Bax-eq} and treat them as $c-$numbers. If one
identifies the Bloch-Floquet factor in \re{scsl2m} as $w(x)=x^N\exp(p)$, then it
is easy to see that the l.h.s.\ of \re{Bax-eq} coincides with the equation of
spectral curve \re{scsl2m}. Going back from the classical to quantum system, one
concludes that the Baxter equation \re{Bax-eq} arises in  result of the
``quantization'' of the spectral curve \re{scsl2m} on the space of the function
$Q(x,\bar x)$. In other words, the constraint imposed by the spectral curve on
the phase space of the classical system becomes an operator equation on the wave
function of the same system after quantization. Thus the Riemann surface
naturally enters into the quantum spectral problem. The important difference with
respect to the classical case is that now the integrals of motion defining the
moduli of the complex structure on \re{scsl2m} must be quantized.

The Baxter equations do not fix the function $Q(x,\bar x)$ uniquely as their
solutions are defined up to multiplication by an arbitrary periodic function
$f(x+i,\bar x)=f(x,\bar x+i)=f(x,\bar x)$. To avoid this ambiguity one has to
impose additional conditions on the solutions to \re{Bax1} and \re{Bax2}. As was
demonstrated in \cite{KKM,kornew}, these conditions follow from the explicit
expression for the kernel of the unitary transformation \re{Phi-def} and they can
be formulated as requirement for $Q(x,\bar x)$ to have correct analytical
properties and asymptotic behaviour at infinity. Namely, choosing
$x=\lambda-in/2$ and $\bar x=\lambda+in/2$ in accordance with \re{x-quan}, and
extending the values of $\lambda$ from real axis to the complex plane, one
requires that $Q_{q,\bar q}(x,\bar x)$ should be a meromorphic function of
$\lambda$ for fixed integer $n$ with an infinite set of poles of the order not
higher than $N$ situated at the points
\be
\{x_{m}^{\pm}=\mp im\,,\ \bar x_{\bar m}^{\pm}=\mp
i(\bar m-1)\}\,,\qquad m,\bar m=1,2,...
\label{poles}
\ee
with $m$ and $\bar m$ being positive integer. In addition, the asymptotic
behaviour of the solutions to the Baxter equation at infinity is given by
\be
Q(\lambda-in/2,\lambda+in/2)\stackrel{\lambda\to\infty}{\sim}
\e^{i\Theta}\lambda^{h+\bar h-N}+\e^{-i\Theta}\lambda^{1-h+1-\bar h-N}
\label{Q-asym}
\ee
with the phase $\Theta=\Theta_h(q,\bar q)$ depending on the quantum numbers of
the state and the total $SL(2,\mathbb{C})$ spins $h$ and $\bar h=1-h^*$ defined
in \re{q2}. Up to an overall normalization, the above two conditions fix uniquely
the solutions to the Baxter equation, and, in addition, they allow us to
determine the spectrum of quantized integrals of motion $q_k$ and $\bar q_k$.

Using the solution to the Baxter equations, we construct the wave function of the
$N-$reggeon states in the separated coordinates \re{prod-Q}. To find the
corresponding value of the energy one has to apply the resulting expression for
the wave function to the Hamiltonian ${\cal H}_N$ transformed to the separated
coordinates. This leads to the following remarkably simple expression for the
energy of the $N-$reggeon state
\be
{\cal H}_N=i\frac{d}{dx} \ln  \lr{x^{2N}
\left[Q(x-i,x)\right]^* Q(x+i,x)}\bigg|_{x=0}.
\label{H-Q}
\ee
The relations \re{Bax1}, \re{Bax2}, \re{poles}, \re{Q-asym} and \re{H-Q} provide
the basis for calculating the spectrum of the $N-$reggeon states.

The Baxter equations can be solved exactly at $N=2$ and their solution is
expressed in terms of the ${}_3F_2-$hypergeometric series. Going though the
calculation of the energy we arrive at
\be
E_2(h)= \frac{\as N_c}{\pi}\epsilon_2 =  -2\frac{\as N_c}{\pi}
\Re\left[\psi\lr{\frac{1+|n|}2+i\nu}-\psi(1)\right]\,.
\ee
This relation coincides with the well--known expression \ci{bfkl} for the energy
of the $N=2$ Reggeon compound state. The maximum value of the energy
\be
E_2^{\rm max}=\frac{\as N_c}{\pi}4\ln 2
\label{E2-max}
\ee
defines the intercept of the BFKL Pomeron. For $N=3$ the solution to the
Schr\"odinger equation has been first found in \cite{janik2} by direct
diagonalization of the integral of motion $q_3$ in the original $\vec
z-$coordinates. Recently, the same result was obtained within the Baxter equation
approach in \cite{devega,kornew}. The ground state of the $N=3$ system defines
the Odderon state in QCD \cite{LN} and its energy is given by
\be
E_3^{\rm max}=\frac{\as N_c}{\pi}\epsilon_3 =  -\frac{\as N_c}{\pi}\cdot 0.247170...
\ee
Contrary to \re{E2-max}, the energy of the $N=3$ state is negative. This implies
that its contribution to the scattering amplitude \re{amp} decreases at
high-energy as $s\to\infty$, or $x\to 0$. Therefore, it becomes of most interest
to find the solution of the Baxter equations, Eqs.~\re{Bax1} and \re{Bax2}, for
higher $N\ge 4$ Reggeon states and calculate the spectrum of the energies
$\varepsilon_N$ governing the high-energy asymptotics in \re{amp}. This problem
has been recently solved in \cite{KKM,kornew} and the results for $\varepsilon_N$
are shown in the Figure~2. The detailed description of the spectrum can be found
in~\cite{kornew}.

\psfrag{-E_N/4}[cc][cc]{$~~~\varepsilon_N$}
\psfrag{N}[cc][cc]{$N$}
\begin{figure}[h]
\centerline{\epsfysize6.0cm \epsfbox{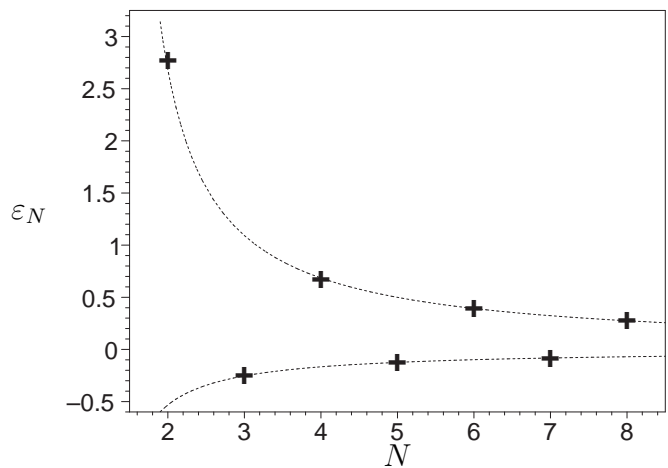}}
\caption[]{The dependence of the energy of the $N-$reggeon states,
$E_N^{\rm max}={\as N_c}\varepsilon_N/{\pi}$, on the number of particles $N$. The
exact values of the energy are denoted by crosses. The upper and the lower dashed
curves stand for the functions $1.8402/(N-1.3143)$ and $-2.0594/(N-1.0877)$,
respectively.}
\label{Fig:E_N}
\end{figure}

It was found in \cite{kornew} that the $N-$reggeon states have different
properties for even and odd $N$. For even $N$, the ground state of the system is
{\it unique\/}, whereas for odd $N$ it is {\it double degenerate\/}. In the
latter case, the color-singlet compound states built odd number of reggeized
gluons have the same energy but different quantum numbers with respect to the
charge conjugation. For odd $N$ the $N-$reggeon states with the charge parity
$C=1$ and $C=-1$ belong to the Pomeron and the Odderon sector, respectively. For
even $N$, the reggeon states have the charge parity $C=1$.

For {\it odd\/} $N$ the ground state energy, $\varepsilon_N$, is {\it negative\/}
and it increases with $N$ approaching the value $\varepsilon_{2\infty+1}=0$ from
below. For {\it even\/} $N$, the ground state energy is {\it positive\/} and it
decreases with $N$ approaching the same value $\varepsilon_{2\infty}=0$ from
above. We recall that for $\varepsilon_N>0$ ($\varepsilon_N<0$) the contribution
of the $N-$reggeon state to the cross-section \re{amp} increases (decreases) at
high energy $s$. Thus, in the Pomeron and the Odderon sectors, the contribution
of the $N-$reggeon states to the cross-section ceases to depend on the energy $s$
as $N\to\infty$. It is interesting to notice that this result has been
anticipated a long time ago within the bootstrap approach \cite{Ven}.

As we have seen in the previous Section, the classical analogs of the $N-$reggeon
states are described by the finite-gap soliton waves propagating in the system of
$N$ particles on the plane. The solitons are uniquely specified by the Riemann
surface \re{Gamma_N} whose moduli are defined by the conserved charges $q_k$. The
solution to the Schr\"odinger equation for the $N-$reggeon states leads to the
quantization of these charges. This allows us to interpret the whole quantization
procedure described in this Section as quantization of the moduli space of the
Riemann surface \re{Gamma_N}.

\section{Universality class of the Regge limit}

In the previous Section the Riemann surfaces appeared within the context of
high-energy QCD as auxiliary objects which were introduced to solve the equations
of motion of the classical system of $N$ reggeized gluons and the Baxter
equations for the corresponding quantum system. In this Section we shall argue
that these surfaces have a definite physical meaning. Actually the situation is
parallel to SUSY YM case where the auxiliary Riemann surface later on was
identified as a part of the six-dimensional worldvolume of the M5 brane. The
remaining four dimensions provide the worldvolume of the theory under
consideration.

Here we shall develop a similar brane picture for the Regge limit of multi-colour
QCD. To do this we explore the analogy with the $\cal{N}$=2 SUSY YM where the
stringy/brane picture naturally emerges from the hidden integrability governing
the low-energy effective actions. We shall argue that in the Regge limit the
Riemann surface which has been considered earlier plays a similar role providing
the worldvolume to the M2 brane describing the scattering process. In this way
this Riemann surface fixes the universality class of the multi-colour QCD in the
Regge limit.


\subsection{Riemann surfaces and QCD versus SUSY YM}

Let us recall the main features of the low-energy effective action in the
$\cal{N}$=2 SUSY YM theories relevant for our purposes \cite{sw1,sw2}. The key
point is that the theory has the nontrivial vacuum manifold since the potential
involves the term
\beq
V(\phi)= \Tr [\phi,\phi^{+}]^2\,.
\eeq
Here $\phi$ is the complex scalar field which generically develops the vacuum
expectation value
\beq
\phi={\rm diag}(\phi_1,....,\phi_{N_c})\,,
\eeq
with $\Tr \phi=0$. The gauge invariant order parameters $u_k=\vev{0|\Tr\phi
^k|0}$ parameterize the Coulomb branch of the vacuum manifold. They define a
scale in the theory with respect to which one can discuss the issue of a low
energy effective action. This action takes into account one-loop perturbative
correction as well as the whole instanton series and it is governed by the
Riemann surface of the genus $N_c-1$ for $SU(N_c)$ gauge group. The same Riemann
surface appears as the spectral curve of a classical integrable many-body system
(see \cite{gm} for a review) which turns out to be the Calogero type systems or
some spin chains. The period matrix $\tau_{ij}(u_k)$ of this Riemann surface
yields the effective coupling constants of the gauge theory.

For example, in the $SU(2)$ case one has
\beq
\tau_{\rm eff}(u_2)= i\frac{4\pi}{g^2(u_2)} +  \frac{\theta(u_2)}{2\pi}
\label{tau}
\eeq
where $g^2(u_2)$ and $\theta(u_2)$ are the effective coupling and $\theta-$term
respectively. The spectrum of the BPS states in this theory is given by
\beq
M_{nm}=|na(u_2) +ma_{D}(u_2)|
\label{BPS-spectr}
\eeq
where $a$ and $a_D$ are the periods of some meromorphic differential
$\lambda_{_{\rm SW}}$ on the spectral curve $\Sigma$
\beq
a=\oint_{A} \lambda_{_{\rm SW}} ,\qquad a_{D}= \oint _{B} \lambda_{_{\rm SW}}\,,
\label{periods}
\eeq
which in this case is a torus
\be
\Sigma_{SU(2)}: \qquad y^2=(x^2-\Lambda^4)(x-{u_2})\,.
\label{Sigma}
\ee
The prepotential $\cal{F}$ defining the low energy effective action is determined
by the relations
\beq
\tau_{\rm eff}(u_2)= \frac{\partial a_D}{\partial a}= \frac{\partial^2 \cal{F}}{\partial^2 a}\,,\qquad
a_D= \frac{\partial {\cal F}}{\partial a}
\eeq
with $\tau_{\rm eff}(u_2)$ given by \re{tau}.

The connection with integrable system emerges when one studies the dependence of
the prepotential on the fundamental scale $\Lambda$
\beq
\frac{\partial {\cal F}(u_2)}{\partial \ln \Lambda}=\beta u_2\equiv \beta H\,,
\eeq
where $\beta$ is the one-loop beta-function of the gauge theory. This equation
has another interpretation as evolution equation of an integrable system where
$H$ coincides with the Hamiltonian of this system and $\ln\Lambda$ is the
evolution time variable. The integrable system provides the natural explanation
for the appearance of the meromorphic differential $\lambda_{\rm SW}$, which
turns out to be the action differential in the separated variables 
 $\lambda_{_{\rm SW}}= p\,dx$. Let us emphasize that  for SUSY
YM case the $\it{classical}$ integrable system is relevant  and the meaning of
the quantum system and the corresponding spectrum for SUSY YM remains an open
question. It should involve the quantization of the vacuum moduli space  since
the hamiltonian in the dynamical system is nothing but $H = u_2 =\vev{\Tr\Phi
^2}$. 
Simultaneously, this parameter serves as the coordinate on the moduli space
of the complex structures of the Riemann surfaces whiich means that
 the quantization of the
integrable system is related with the 
quantization of the effective d=2 gravity.

The Riemann surface $\Sigma$ degenerates at some points on the Coulumb branch 
 on the moduli space. After the soft breaking of $\cal{N}$=2 SUSY down to $\cal{N}$=1
SUSY these points correspond to the vacuum states of the $\cal{N}$=1 gauge
theory. Some massless states condense at these points leading to the formation of
a mass gap and to confinement. In the $SU(2)$ case, Eq.~\re{Sigma}, the
$\cal{N}$=1 vacua correspond to the points $u_2=\pm\Lambda^2$ at which the
monopoles or dyons become massless and condense.

Let us now consider a particular theory, namely the superconformal $\cal{N}$=2
SUSY YM with $N_f=2N_c$ massless fundamental hypermultiplets \cite{sw2,aps,ho}.
The corresponding integrable system is described by the spectral curve
$\Sigma_{N_c}$
\beq
 y^2=P_{N_c}^2(x) -4x^{2N_c}(1- \rho^2(\tau_{cl}))\,,
\label{SUSY-curve}
\eeq
where $\rho^2(\tau_{cl})$ is some function of a coupling constant of the theory
(see Eq.~\re{rho} below) and the polynomial $P_{N_c}$ depends on the coordinates
on the moduli space $\vec u=(u_2,...,u_{N_c})$
\be
P_{N_c}(x)= \sum_{k=0}^{N_c} q_k(\vec u)\, x^{N_c-k} =2x^{N_c}+ q_2\, x^{N_c-2} +
... + q_{N_c}\,,
\ee
where $q_0=2$, $q_1=0$ and other $q_k$ are some known functions of $\vec u$.
Their explicit form is not important for our purposes. The Seiberg-Witten
meromorphic differential on the curve \re{SUSY-curve} is given by
\be
\lambda_{_{\rm SW}}= p\,dx= \ln({\omega}/{x^{N_c}})\,dx\,,
\label{lambda-SW}
\ee
where $y=\omega-x^{2N_c}/\omega$.

{}From the point of view of integrable models the spectral curve \re{SUSY-curve}
corresponds to a classical XXX Heisenberg spin chain of length $N_c$ with the
spin zero at all sites (due to $q_1=0$) and parameter $\rho$ related to the
external magnetic field, or equivalently, to the twisted boundary conditions
\cite{ggm}.

Let us compare the spectral curve \re{SUSY-curve} for the superconformal
$\cal{N}$=2 SUSY YM with $N_f=2N_c$ with the spectral curve \re{Gamma_N} for
$N-$reggeon compound states in multi-colour QCD. It is amusing to observe that
these two curves coincide if we make the following identification
\begin{itemize}
\item The number of the reggeons $N=N_c$;
\item The integrals of motion of multi-reggeon system are identified as the above
mentioned functions $q_k(\vec u)$ on the moduli space of the superconformal
theory;
\item The coupling constant of the gauge theory should be such that
$\rho(\tau_{cl})=0$.
\end{itemize}
Under these three conditions the both theories fall into the same universality
class.

Let us clarify the meaning of the last condition. Since the superconformal theory
has a vanishing $\beta-$function one can assign a definite value to the bare
coupling constant $\tau_{\rm cl}$. The function $\rho(\tau)$ entering
\re{SUSY-curve} at $\tau=\tau_{\rm cl}$ is given by some combination of the
modular forms~\cite{minahan,argyres} which can be determined from the duality
properties of the superconformal $SU(N_c)$ gauge theory. It turns out that the
duality groups are different for odd and even $N_c$. The group for even $N_c$ is
generated by
\beq
T:\quad\tau \rightarrow \tau +1, \qquad S:\quad\tau \rightarrow -1/\tau\,.
\eeq
The group for odd $N_c$ has the same $T-$transformation and
\beq
S:\quad\tau \rightarrow -\frac{1}{(2\cos(\pi/2N_c))^2\tau}
\eeq
subject to the constaints
\beq
S^2=1 \qquad (ST^{-1})^{2N_c} =1
\eeq
The explicit formulae for $\rho(\tau)$ can be expressed in terms of weight four
automorphic forms \cite{minahan} for the duality group
\beq
\rho(\tau)=\frac{G_4(\tau) + H_4(\tau)}{G_4(\tau) - H_4(\tau)}\,,
\label{rho}
\eeq
where $G_4(\tau)$ is the Poincare series
\beq
G_4(\tau) = \sum_{c,d}(c\tau + d)^{-4}\,,\qquad
H_4(\tau)=\frac1{\alpha^2\tau^4}G_4\left(-\frac1{\alpha\tau}\right)\,,
\eeq
with $\alpha=1$ for even $N_c$ and $\alpha=4\cos^2(\pi/(2N_c))$ for odd $N_c$.
Here the summation goes over all pairs of integers $(c,d)$ which occur in modular
transformations $\tau\rightarrow (a\tau + b)/(c \tau + d)$ generated by $T$ and
$STS^{-1}$. Finally, using \re{rho} one finds that the condition
$\rho(\tau_{cl})=0$ is satisfied for
\beq
\tau_{\rm cl}= \frac{1}{2}+ \frac{i}{2}\tan \frac{\pi}{2N_c}\,.
\label{tau-QCD}
\eeq

The moduli of this superconformal theory are defined by the vector of the scalar
condensates $\vec u$ and bare coupling constant $\tau_{\rm cl}$. The spectral
curve \re{SUSY-curve} degenerates when $\rho\to\infty$ or when the discriminant
of the curve is zero. Let us consider simplest nontrivial case $N_c=3$ which on
the QCD side corresponds to the Odderon ($N=3$) state. In this case the
discriminant is given by
\be
(1-f)f^{3}\left[f -\left(1+ \frac{2q_2^{3}}{27q_3^{2}}\right)^2\right]q_3^{10}=0
\ee
where $f=1-\rho^2$. The curve degenerates at four values of $\rho^2$ among which
three $\rho^2=\infty$, $0$ and $1$ are universal and the last one depends on the
moduli $q_2$ and $q_3$ on the Coulomb branch. The first one is a strong coupling
point corresponding to $\tau_{\rm cl}=0$. The second one is the so-called strong
coupling orbifold point in which $\tau_{\rm cl}$ is given by \re{tau-QCD}. It is
this point on the moduli space which corresponds to the Regge limit of
multi-colour QCD. Finally, the third one is a weak coupling point $\tau_{\rm
cl}\to\infty$. For $N_c=3$ the explicit expression for the function $\rho(\tau)$,
Eq.~\re{rho}, is given in terms of the Dedekind $\eta-$function by~\cite{minahan}
\be
\rho(\tau)= \frac{f_{+}(\tau)}{f_{-}(\tau)},\qquad
f_{\pm}(\tau)=\lr{\frac{\eta^3(\tau)}{\eta(3\tau)}}^3 \pm
\lr{3\frac{\eta^3(3\tau)}{\eta(\tau)}}^3
\ee
The strong coupling orbifold point $\rho(\tau_{\rm cl})=0$ describing the Odderon
state in QCD occurs at
\beq
\tau_{\rm cl}=\frac12 +\frac{i}{2\sqrt{3}}\,.
\label{tau-N=3}
\eeq
Let us note this point becomes the Argyres-Douglas point when the moduli
dependent singularity of the spectral curve $\rho^2(\tau)=1-\left[1+
{2q_2^{3}}/(27q_3^{2})\right]^2$ collides with the strong coupling orbifold point
\re{tau-N=3}. This happens when either $q_2/q_3=0$, or
\be
q_3=\pm \frac{(-q_2)^{3/2}}{\sqrt{27}}\,.
\label{AD-point}
\ee

Going over to the general case of the $SU(N_c)$ gauge theory one
finds~\cite{minahan,argyres} that independently on $\vec u$, the spectral curve
\re{SUSY-curve} has three universal singularities at $\rho^2(\tau_{\rm
cl})=\infty$, $0$ and $1$ which have the same interpretation as in the case
$N_c=3$.

Finally we would like to note that there is the following intriguing fact. In the
case of $N_c=2$ which corresponds on the QCD side to the BFKL Pomeron state, the
effective coupling constant is given in the weak coupling regime by the
expression
\be
\tau_{\rm eff}= \tau_{\rm cl} + i\frac{4\ln 2}{\pi}
+ \sum_{k}c_k\cdot \e^{2i\pi k\tau_{\rm cl}}, ~~~
\ee
where the second term  is due to a finite one-loop
correction~\cite{khoze} and the rest is the sum of instanton contributions. It is
amusing that this one-loop correction to the coupling constant
\be
\frac{1}{g^2_{\rm eff}}=
\frac{1}{g^2_{\rm cl}}\left[1 + \frac{g^2_{\rm cl}}{4\pi^2} 4\ln 2\right] +...
\ee
 coincides with the expression for the
intercept of the BFKL Pomeron \re{E2-max} after one identifies bare coupling
constant in the superconformal 
theory with the t'Hooft coupling constant in QCD.

It would be very interesting to understand a proper QCD interpretation of the
nonzero values of $\rho$ in \re{SUSY-curve}. From the point of view of spin
chains it corresponds to the external magnetic field, or equivalently to the
twisted boundary conditions \cite{ggm}. The natural conjecture for $\rho\neq 0$
to occur is to consider the compound states in multi-colour QCD in which one of
the reggeized gluons is replaced by the pair of reggeized quark and
antiquark~\cite{Kirschner}. Similar picture takes place in the evolution
equations for three-quark baryonic light-cone operators which will be discussed
in Section 6. In that case we shall deal with a real $SL(2,\mathbb{R})$ spin
chain whose spectrum contains special eigenstates (with $q=0$ in Eq.~\re{Q}
below) for which a pair of quarks effectively behaves as one spin site~\cite{bm}.
For generic values of $\rho$ we can have a genus one curve even for a two
particle system and the whole power of the duality group can be used. For
example, the behaviour of the system near the strong coupling orbifold point
$\rho=0$ can be related to the dual system near the weak coupling point $\rho=1$.
Certainly this possibility of having $\rho\neq 0$ in QCD deserves further
investigation.

\subsection{Brane picture for the Regge limit}

Let us turn now to a stringy/brane picture for the Regge limit in multi-colour
QCD.

To warm up we would like to recall the brane description of the low-energy
dynamics of the ${\cal N}=2$ SUSY YM. In the IIA framework the pure gauge theory
is defined on the worldvolume of $N_c$ D4 branes with the coordinates
$(x_0,x_1,x_2,x_3,x_6)$ stretched between two NS5 branes with the coordinates
$(x_0,x_1,x_2,x_3,x_4,x_5)$ and displaced along the coordinate $x_6$ by an amount
inversely proportional to the coupling constant, $\delta x_6=1/g^2$
\cite{witten}. The coordinates at which the D4 branes intersect with the
$(x_4,x_5)$ complex plane define the vacuum expectation value of the scalar
fields. This picture agrees perfectly with the RG behaviour of the coupling
constant and yields the correct beta function in the gauge theory. The Riemann
surface $\Sigma$ discussed above describes the vacuum state of the theory and the
spectrum of the stable BPS states. The lifting to the M theory picture leads to
emergence of a single M5 brane with the worldvolume $R^4\times\Sigma$.

In our case, we also need to incorporate into this picture branes corresponding
to the fundamental matter with $N_f=2N_c$. There are two ways to do this: either
using semi-infinite D4 branes lifted into M5 brane in the M theory, or using D6
branes which induce  the nontrivial KK monopole background for the M2
brane wrapped on  the Riemann surface\cite{witten}.
As was shown in \cite{ggm} in the latter case the resulting
brane picture remains consistent with the integrable spin chain dynamics and we
shall stick to this case.

The explicit metric of the KK background in the M theory involving
$(x_4,x_5,x_6,x_{10})$ coordinates has the multi-Taub-NUT form
\beq
ds^2= \frac{V}{4}d\vec{r}^2 + \frac{V^{-1}}{4}(d\tau +\vec{A}d\vec{r})^2
\eeq
where $\vec{r}=(x_4,x_5,x_6)$,~ $\tau =x_{10}$ and
$\vec{A}$ is the Dirac monopole potential. The magnetic charge
comes  from the nontrivial twisting of $S^1$ bundle over $R^3$.
The function $V$ behaves as
\beq
V= 1 + \sum_{i=1}^{i=N_f} \frac{1}{|\vec{r} - \vec {r_i}|}
\eeq    
where $\vec r_i = (x^i_4,x^i_5,x^i_6)$ are the positions of six-branes. For a
superconformal case one must have  $x^i_4 = x^i_5 =0$ and positions of sixbranes
in $x_6$ direction are irrelevant.

Finally let us remind the brane interpretation of the BPS spectrum in the ${\cal
N}=2$ SUSY YM. One way to realize these states is to consider the self-dual
noncritical strings (which come from M2 branes ending on the surface) wrapped
around the Riemann surface \cite{vafa}. Since one has to keep some amount of SUSY
the wrapping of the string should be along geodesics in some metric. This is
equivalent to looking for the geodesics on $\omega$ plane (in IIB
approach), where $\omega$ is the same as in Eq.(\ref{scsl2m}) and  is
related to $y = \exp(-x_6 + ix_5)$ and $x = x_4+ix_5$ as $y =\omega -
 x^{2N_c}/\omega$.
The metric on $\omega$ plane  is defined in
terms of Seiberg-Witten differential
\beq
ds^2 =\lambda_{_{\rm SW}} \bar{\lambda}_{_{\rm SW}}
\eeq
where $\bar{\lambda}_{_{\rm SW}}=\bar p\,d\bar x=(\lambda_{_{\rm SW}})^*$. Closed
geodesics in the $\omega-$plane can be immediately lifted to closed curves on the
Riemann surface. Hence the existence of the BPS state is the related to the
existence of the closed geodesics in the $(n,m)-$cohomology class. Another
important realization of BPS states arises from IIB/F theory picture. Namely the
$(n,m)$ BPS states come from the $(n,m)$ strings which are stretched between the
origin (which is generically split nonperturbatively into two strong coupling
singularities) and the position of D3 brane we are living on~\cite{sen}. Now the
nontrivial metric on the Coulomb branch of the moduli space is involved. It has
the form of the cosmic string metric with the conical singularity which can be
brought into the flat form on the  cover. In this picture $SL(2,\mathbb{Z})$
invariance of the spectrum follows from the invariance of the IIB string theory.


Let us turn now to our proposal for
the brane realization of the Regge limit. We
shall explore the brane representations for the $N_f=2N_c$ theory known in the
IIA/M theory \cite{witten} and IIB/F theory \cite{sen}. However unlike the SUSY
case where the spectral curve is
embedded in the internal ``momentum'' space the
spectral curve of the noncompact spin chain is placed in the phase space
involving both impact parameter plane as well as momenta.

We shall first consider IIA/M type picture which is reminiscent to the
realization of SYM theory via two NS5 and $N_c$ D4 branes. We suggest that the
coordinates involved in ``IIA'' picture are the transverse impact parameter
coordinates $x_1,x_2$ and rapidity $\lambda=\ln k_{+}/k_{-}$. Transverse
coordinates are analogue of $(x_4,x_5)$  coordinates in SUSY case while rapidity
substitutes the $x_6$ coordinate.

Now let us make the next step and suggest that similar to SUSY case the single
brane is wrapped around the spectral curve of XXX magnet and two ``hadronic
planes" together with N ``Reggeonic strings'' are just different projections of
the single M2 brane with worldvolume $R\times \Sigma$. The coordinates involved
into configurations are $x_1+i x_2$ and $y=e^{-(\lambda +ix_{10})}$ where
$x_{10}$ is the ``M-theory'' coordinate. More precisely $y$  can be identified
with the Bloch-Floquet factor $\omega$ in the definition of the Baker-Akhiezer
function \re{pbc}.
The correspondence between ``IIA'' and ``M-theory'' picture is shown in Fig.3.

\begin{figure}
\begin{center}
\epsfxsize=5in
\epsfbox{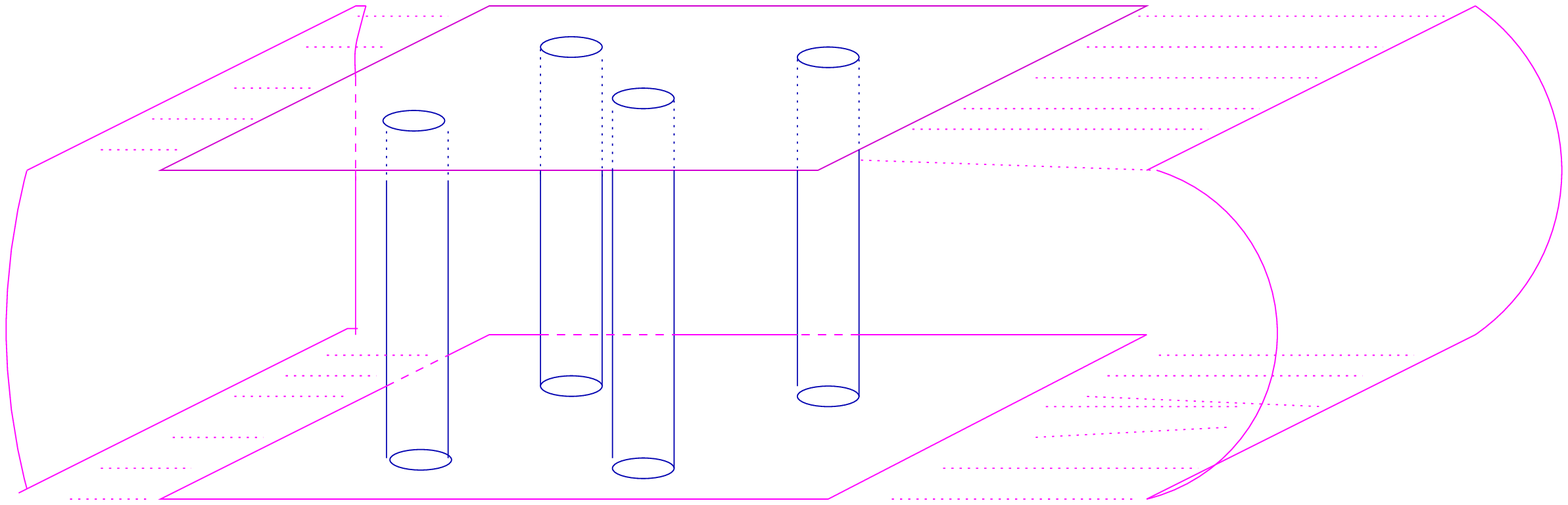}
\caption{Spectral curve}
\end{center}
\end{figure}

Let us emphasize once again that
the brane configuration for Regge limit contrary
to SYM case partially involves
the coordinate space. More precisely the geometry
of $N_f=2N_c$ theory is determined by the following parameter \cite{witten}
\be
\xi = - \frac{4\lambda_{+}\lambda_{-}}{(\lambda_{+}-\lambda_{-})^2}\,.
\ee
Here $\lambda_{+}$ and $\lambda_{-}$ are asymptotic positions of five-branes
defined by the large $x$ behaviour of the curve
\beq
\omega \propto \lambda_{\pm} x^{N_c}\,.
\eeq
$\lambda_\pm$ can be found as roots of the equation
\beq
\lambda_\pm^2 +\lambda_\pm +\frac{1}{4}(1-\rho^2)=0
\eeq
Since Regge limit corresponds to the strong coupling orbifold point, $\rho=0$,
the value of $\xi$ is fixed as $\xi=\infty$. This corresponds to the coinciding
branes at infinity.

Finally, the M theory brane picture for the Regge limit involves M5 brane
corresponding to the vacuum state of the QCD. We can not say how it is placed
precisely as the minimal surface in the internal seven dimensional space since
the corresponding geometry is unknown yet. The new ingredient - M2 brane share
the time direction with the M5 brane and wrapping around the Riemann surface
which is embedded into two-dimensional complex ``phase space'' with the
multi-Taub-NUT metric determined by KK monopoles with the magnetic charge $2N$,
which is a double number of reggeized gluons participating in the scattering
process.


\section{Quantum spectrum and S-duality}

The $S-$duality is a powerful symmetry in the SUSY YM theory which allows us to
connect the weak and strong coupling regimes. The effective coupling in this
theory coincides with the modular parameter of the spectral curve of the
underlying classical integrable model. As a consequence, the $S-$duality
transformations in the gauge theory are translated into the modular
transformations of the spectral curve describing complexified integrable system.
A formulation of the $S-$duality in the latter system naturally leads to an
introduction of the notion of the dual action~\cite{fgnr}.

So far the $S-$duality was well understood only for classical integrable models.
In the case of multi-colour QCD in the Regge limit the situation is more
complicated since the duality has to be formulated for a quantum integrable
model. The integrals of motion take quantized set of values and the coordinates
on the moduli space are not continuous anymore. Therefore the question to be
answered is whether it is possible to formulate some duality transformations at
the quantum level.

To study this question let us propose the WKB quantization conditions which are
consistent with the duality properties of the complexified dynamical system whose
solution to the classical equations of motion are described by the Riemann
surface. We recall that the standard WKB quantization conditions involve the real
slices of the spectral curve
\beq
\oint_{A_i}p\,dx = 2\pi \hbar (n_i+1/2)
\eeq
where $n_i$ are integers and the cycles $A_i$ correspond to classically allowed
trajectories on the phase space of the system. In our case the coordinate $x$ is
complex and arbitary point on the Riemann surface is classically allowed. As a
result the general classical motion involves both $A-$ and $B-$cycles on the
Riemann surface. This leads to the following generalized WKB quantization
conditions (see also \cite{kornew})
\beq
\Re\oint_{A_i}p\,dx = \pi \hbar n_i\,, \qquad \Re\oint_{B_i}p\,dx = \pi\hbar
m_i\,,
\label{WKB-new}
\eeq
where the ``action'' differential was defined in \re{lambda-SW}. Note that in the
context of the SUSY YM this condition would correspond to the nontrivial
constraints on the periods \re{periods} and on the mass spectrum of the BPS
particles \re{BPS-spectr}. It is clear that the WKB conditions \re{WKB-new} imply
the duality $A_i\leftrightarrow B_i$ and $n_i\leftrightarrow m_i$.

The conditions \re{WKB-new} are analogous to the WKB quantization for a particle
in a multi-dimensional case. We have exactly this situation since we are dealing
with complexified phase space. For a particle in higher dimensions one considers
the representation of the wave function in terms of the multivalued functions
$A_k$ and $S_k$ on the $\vec x$-space
\beq
\Psi(\vec x) =\sum_k A_k(\vec x)\exp\lr{\frac{i}{\hbar}S_k(\vec x)}\,.
\eeq
Here $A_k(\vec x)$ and $S_k(\vec x)$ are different branches of the multivalued
functions $A(\vec x)$ and $S(\vec x)$. To obtain the WKB quantization conditions
one considers the covering space on which $A(\vec x)$ and $S(\vec x)$ are single
valued
and requires that the wave function $\Psi(\vec x)$ should resume its original
value after encircling along all noncontructable cycles $C_i$ on the covering
manifold. This leads to
\beq
\oint_{C_i}\vec p\cdot d\vec x + \oint_{C_i} d \ln A(\vec x) = 2\pi \hbar n_i\,.
\eeq
The second integral is equal to the sum of two terms related to the topological
invariants of the classical trajectory: the number of intersections of $C_i$ with
hypersurface of the caustics and the number of intersections of $C_i$ with the
hypersurface of the turning points. In our case $\oint_{C_i}\vec p\cdot d\vec
x=2\Re\oint_{C_i}p\,dx$ and $\oint_{C_i} d \ln A(\vec x)$ is an even integer.

Let us consider the quantization conditions \re{WKB-new} in the simple case of
the Odderon $N=3$ system. The spectral curve \re{Gamma_N} is a torus
\beq
y^2=(2x^3+q_2x +q_3)^2 - 4x^6
\eeq
where $q_2$ is given by \re{q2} and \re{h} while $q_3$ is the complex integral of
motion to be quantized. The quantization conditions \re{WKB-new} read (for $\hbar
=1$)
\beq
\Re a(q_3) = \pi n, \qquad \Re a_D(q_3) = \pi m
\label{a-N=3}
\eeq
where $n$ and $m$ are integer. These equations can be solved for large values of
$q_3^2/q_2^3 \gg 1$, for which the expressions for the periods $a(q_3)$ and
$a_{D}(q_3)$ are simplified considerably. The explicit evaluation of the
integrals \re{periods} in this limit  yields
\be
a(q_3) = \frac{(2\pi)^2 q_3^{1/3}}{\Gamma^3(2/3)}\,,\qquad a_D(q_3)
=a(q_3)\lr{\frac12+\frac{i}{2\sqrt{3}}}
\ee
Substituting these expressions into \re{a-N=3} one finds~\cite{kornew}
\beq
q_3^{1/3}=\frac{\Gamma^3(2/3)}{2\pi}\left(\frac{\ell_1}2+i\frac{\sqrt{3}}2
\ell_2\right)\,.
\label{q3-quan}
\eeq
where $\ell_1=n$ and $\ell_2=n-2m$. The same quantization conditions
 can be also written as
\be
(\ell_1+3\ell_2)\cdot a(q_3)-6\ell_2\cdot a_{D}(q_3) =\pi (\ell_1^2+3\ell_2^2)\,.
\ee
The WKB expressions \re{q3-quan} are in a good agreement with the exact
expressions for quantized $q_3$ obtained from the solutions of the Baxter
equations in \cite{kornew}.

There is a simple relation between the modular parameter of the curve with the
periods $a$ and $a_D$
\beq
a_{D}= \tau_{\rm eff}\,a\,.
\label{W}
\eeq
It follows from the fact that $a$ and $a_D$ are the solutions to the Picard-Fuchs
equations which in the Odderon case is the second order differential equation.
One can show that the Wronsian for this equation $a a_D'-a_D a'$ vanishes and
using $\tau_{\rm eff}=a_D'/a'$ one gets \re{W}. Due to this fact half of
quantization conditions can be formulated in terms $\tau_{\rm eff}$ instead of
$a_D$.

Let us emphasize that the point at the moduli space corresponding to the
degeneration of the torus for the Odderon case does not appear in the quantum
spectrum.  From the point of view of the gauge theory this means that the
appearance of the massless states is forbidden.

In the general multi-reggeon case we have to consider the quantization conditions
\re{WKB-new} on the Riemann surface \re{Gamma_N} of the genus $(N-2)$ which has
the same number of the $A-$ and $B-$cycles. In result the spectrum of the
integrals of motion $q_3$, $...$, $q_N$ is parametrized by two $(N-2)-$component
vectors $\vec{n}$ and $\vec{m}$. In the SUSY YM case these vectors define the
electric and magnetic charges of the BPS states. In the Regge case the physical
interpretation of $\vec{n}$ and $\vec{m}$ is much less evident. Let us first
compare the electric quantum numbers in the two cases. In Regge case it
corresponds to rotation in the coordinate space around the ends of the Reggeons.
This picture fits perfectly with the interpretation of the electric charge in
SUSY YM case. Indeed  VEVs
 of the complex scalar take values on  the  complex  plane which is the 
counterpart of the impact parameter plane and
the rotation of the phase of the complex scalar is indeed the ``electric
rotation''.

To get some guess concerning the ``magnetic'' quantum numbers it is  
instructive to check the geometrical picture behind them in the 
simplified ``IIA'' picture.
 All  states corresponding to the ``electric'' degrees of freedom 
are related to  fundamental strings  encircling  ``reggeoinic'' tubes  
(see Fig.3) and  don't feel the
hadronic planes. However the ``magnetic'' states as is well known from SUSY YM
case are represented by the membrane stretched between two strings and two
hadronic planes. 
Therefore  these states are sensitive to  hadronic  quantum numbers.
More detailed  interpretation of ``magnetic'' degrees of freedom in
the Regge regime  has to be recovered.

Let us emphasize that due to the stringy realization of the
abovementioned  BPS spectrum  we obtain rather transparent picture of
the WKB quantization in the stringy
terms. Remind that, in SUSY YM
 the mass of BPS state $M_{n,m}$ is given by  the energy of
the dyonic $(n,m)$  string stretched between the  value 
of $u_2$ corresponding to a given VEV of a scalar field 
and the point  $u_2=0 $  where orientifold 
and all D7 branes are placed. Hence, the
WKB quantization at least in the Odderon case can be formulated as the
quantization of the energy of the dyonic strings at the complex $q_3$ plane and
the spectrum inherited the duality from S-duality of IIB string.

\section{Stringy/brane picture and the calculation
of the anomalous dimensions}

In the previous Sections we have demonstrated that integrability properties of
the Schr\"odinger equation for the compound state of Reggeized gluons give rise
to the stringy/brane picture for the Regge limit in multi-colour QCD. There is
another limit in which QCD exhibit remarkable properties of integrability. It has
to do with the scaling dependence of the structure functions of deep inelastic
scattering and hadronic light-cone wave functions in QCD. In the both cases, the
problem can be studied using the Operator Product Expansion and it can be
reformulated as a problem of calculating the anomalous dimensions of the
composite operators of a definite twist. The operators of the lowest twist have
the following general form
\ba
&&{\cal O}^{(2)}_{N,k} (0) = (y D)^k \Phi_1(0) (y D)^{N-k}
\Phi_2 (0),
\nonumber
\\[2mm]
&&{\cal O}^{(3)}_{N,\mybf{k}}(0) =(y D)^{k_1} \Phi_1(0) (y D)^{k_2}
\Phi_2(0)(y D)^{N-k_1-k_2}\Phi_3(0),
\ea
where $\Mybf{k}\equiv (k_1,k_2)$ denotes the set of integer indices $k_i$,
$y_\mu$ is a light-cone vector such that $y_\mu^2=0$. $\Phi_k$ denotes elementary
fields in the underlying gauge theory and $D_\mu =
\partial_\mu -i A_\mu$ is a covariant derivative. The operators of a definite
twist mix under renormalization with each other. In order to find their scaling
dependence one has to diagonalize the corresponding matrix of the anomalous
dimension and construct linear combination of such operators, the so-called
conformal operators~\cite{collineargroup,BFLK}
\be
{\cal O}^{\rm conf}_{N,q}(0) = \sum_{\mybf{k}} C_{\mybf{k,q}}\cdot {\cal
O}_{N,\mybf{k}}(0)\,.
\label{conf-op}
\ee
A unique feature of these operators is that they have an autonomous RG evolution
\be
\Lambda^2\frac{d}{d \Lambda^2}\,{\cal O}^{\rm conf}_{N,q}(0)
=-\gamma_{N,q} \cdot{\cal O}^{\rm conf}_{N,q}(0)\,.
\ee
Here $\Lambda^2$ is a UV cut-off and $\gamma_{N,q}$ is the corresponding
anomalous dimension depending on some set of quantum numbers $\Mybf{q}$ to be
specified below. It turns out that the problem of calculating the spectrum of the
anomalous dimensions $\gamma_{N,q}$ to one-loop accuracy becomes equivalent to
solving the Schr\"odinger equation for the $SL(2,\mathbb{R})$ Heisenberg spin
magnet. The number of sites in the magnet is equal to the number of fields
entering into the operators under consideration.

To explain this correspondence it becomes convenient to introduce nonlocal
light-cone operators
\be
F(z_1,z_2)= \Phi_1(z_1y) \Phi_2(z_2y)\,,\qquad F(z_1,z_2,z_3)=
\Phi_1(z_1y) \Phi_2(z_2y) \Phi_3(z_3y)\,.
\label{F}
\ee
Here $y_\mu$ is a light-like vector $(y_\mu^2=0)$ defining certain direction on
the light-cone and the scalar variables $z_i$ serve as a coordinates of the
fields along this direction. The fields $\Phi_i(z_i y)$ are transformed under the
gauge transformations and it is tacitly assumed that the gauge invariance of the
nonlocal operators $F(z_i)$ is restored by including the Wilson lines between the
fields in the appropriate (fundamental or adjoint) representation. The conformal
operators appear in the OPE expansion of the nonlocal operators \re{F} for small
$z_1-z_3$ and $z_2-z_3$.

The field operators entering the definition of $F(z_i)$ are located on the
light-cone. This leads to the appearance of the additional light-cone
singularities. They modify the renormalization properties of the nonlocal
light-cone operators \re{F} and lead to nontrivial evolution equations which as
we will show below become related to integrable chain models. We notice that
there exists the following relation between the conformal three-particle
operators \re{conf-op} and the nonlocal operators \re{F}
\be
{\cal O}^{\rm conf}_{N,q}(0)=
\Psi_{N,q}(\partial_{z_1},\partial_{z_2},\partial_{z_3})
F(z_1,z_2,z_3)\bigg|_{z_i=0}\,,
\ee
where $\Psi_{N,q}(x_1,x_2,x_3)$ is a homogenous polynomial in $x_i$ of degree $N$
\be
\Psi_{N,q}(x_1,x_2,x_3)=\sum_{\mybf{k}} C_{\mybf{k,q}} \cdot x_1^{k_1} x_2^{k_2}
x_3^{N-k_1-k_2}
\ee
with the expansion coefficients $C_{\mybf{k,q}}$ defined in \re{conf-op}. Similar
relations hold for the twist-2 operators.
The problem of defining the conformal operators is reduced to finding the
polynomial coefficient functions $\Psi_{N,q}(x_i)$ and the corresponding
anomalous dimensions $\gamma_{N,q}$. Using the renormalization properties 
 of the
nonlocal light-cone operators \re{F} one can show~\cite{bm}, that to 
the one-loop accuracy the QCD evolution equation for the conformal 
operators \re{F} can be rewritten in the form of a Schr\"odinger equation
\be
{\cal H}\cdot \Psi_{N,q}(x_i)
    =  \gamma_{N,q} \Psi_{N,q}(x_i)\,,
\label{gamma}
\ee
where the Hamiltonian ${\cal H}$ acts on the 
$x_i-$variables which are conjugated
to the derivatives $\partial_{z_i}$ and, therefore, have the meaning of
light-cone projection $(y\cdot p_i)$ of the momenta $p_i$ carried by particles
described by fields $\Phi(z_i y)$.

The Hamiltonian ${\cal H}$ has the following remarkable properties.
\begin{itemize}
\item[(i)]
This Hamiltonian is a sum of two-particle Hamiltonians
\end{itemize}
 \be
 {\cal H}^{(2)} = H_{12}
 \,,\qquad
{\cal H}^{(3)} = H_{12} +H_{23}+H_{13}
\ee
where $H_{ij}$ acts on the coordinates of particles $i$ and $j$.
\begin{itemize}
\item[(ii)]
All two-particle Hamiltonians  $H_{ij}$   are invariant under $SL(2,R)$
transformations
\end{itemize}
\be
 [ H_{ij}, \vec S_i+\vec S_j ] = 0
\label{commHS}
\ee
where one-particle $SL(2,R)$ 
generators $\vec S_i$ act on the space of functions
$\Psi(x_i)$ as
\begin{eqnarray}
S_{k,0}\Psi(x_k) &=& \left(x_k\partial_k +j_k\right) \Psi(x_k)\,,\quad
\nonumber\\
S_{k,+}\Psi(x_k) &=& -x_k \Psi(x_k)\,,\quad
\nonumber\\
S_{k,-}\Psi(x_k) &=&
\left(x_k\partial_k^2 +2j_k\partial_k\right)\Psi(x_k).
\label{adjoint}
\end{eqnarray}
Let us note that these generators where $x_k$ play the 
role of momenta variables
are dual to generators \re{spins} where $z_k$ were coordinates.  This symmetry of
one-loop evolution kernels $H_{ij}$ follows from the invariance of classical QCD
Lagrangian under the group of conformal transformations \cite{collineargroup}
which is reduced to its $SL(2,R)$ subgroup on the light-cone
\be
z\to z'=\frac{az+b}{cz+d}\,,\qquad
\Phi_k(zy)  \to \Phi_k'(z'y)=(cz+d)^{-2j_k}\Phi_k\left(\frac{az+b}{cz+d}y\right)
\label{Moebius}
\ee
with  $ad-bc=1$ and $a,b,c,d$ real. Here
\begin{equation}
  j_k=\frac{1}{2}(d_k+s_k),
\label{cspin2}
\end{equation}
where $d_k$ is a canonical dimension of a field $\Phi_k(z y)$ and $s_k$ is a
light-cone component of it spin. For example, $d_q=3/2$ and $s_q = 1/2$ and $j_q
=1$ for quarks. Due to \re{commHS} Hamiltonians $H_{ij}$ are functions of the
two-particle Casimir operators
\begin{equation}
  (\vec S_1+\vec S_2)^2 = J_{12}(J_{12}-1).
  \label{def:J12}
\end{equation}
For example  when $\Phi_1$ and $\Phi_2$ are quark fields of the same chirality
\be
F_{\alpha\beta}(z_1,z_2)=\sum_{i=1}^{N_c} (\bar
q_i^\uparrow{\not\hspace*{-0.59mm}y} )_\alpha(z_1y) ({\not\hspace*{-0.55mm} y}
q_i^\uparrow)_\beta(z_2y)
\ee
with $q_i^\uparrow=(1+\gamma_5)q_i/2$, the two-particle Hamiltonian is given
by~\cite{BFLK}
\begin{eqnarray}
  H_{12} = \frac{\as}{\pi}C_F \left[H_{qq}(J_{12})+1/4\right]\,,\qquad
  H_{qq}(J_{12})= \psi(J_{12})-\psi(2).
\label{H-qq}
\end{eqnarray}
where $C_F=(N_c^2-1)/(2N_c)$. The eigenfunctions for this Hamiltonian are the
highest weights of the discrete series representation of the $SL(2,R)$ group
\be
\Psi_{N}^{(2)}(x_1,x_2) = (x_1+x_2)^N C_N^{3/2} \left(\frac{x_1-x_2}{x_1+x_2}\right)
\ee
where $C_N^{3/2}$ are Gegenbauer polynomials. The corresponding eigenvalues
define the anomalous dimensions of the twist-2 mesonic operators built from two
quarks with the same helicity
\begin{eqnarray}
  \gamma_N^{(2)} = \frac{\as}{\pi}C_F[\psi(N+2)-\psi(2)+1/4]
  = \frac{\as}{\pi}C_F\left[\sum_{k=1}^N\frac 1{k+1}+\frac14\right]\, .
\end{eqnarray}
At large $N$ this expression has well-known asymptotic behaviour
$\gamma_N^{(2)}\sim \as C_F/\pi\ln N$.

It is  conformal symmetry which dictates that the two-particle Hamiltonian is a
function of  the Casimir operator of the $SL(2,\mathbb{R})$ group, but it does
not fix this function. The fact that this function turns out to be the Euler
$\psi$-function  leads to a hidden integrability of the evolution equations for
anomalous dimensions of baryonic operators in  very much the same way as
emergence of the Euler $\psi$-functions in 
the BFKL kernel \re{psi} leads to the
integrability of multi-colour QCD in the Regge limit. Namely, for baryonic
operator built from three quark fields of the same chirality
\be
F_{\alpha\beta\gamma}(z_1,z_2,z_3)=\sum_{i,j,k=1}^{N_c}\epsilon_{ijk}({\not\hspace*{-0.55mm}
y} q_i^\uparrow)_\alpha(z_1y) ({\not\hspace*{-0.55mm} y}
q_j^\uparrow)_\beta(z_2y) ({\not\hspace*{-0.55mm} y} q_k^\uparrow)_\gamma(z_3y)
\ee
the evolution kernel is given by~\cite{BDM,bm}
\be
{\cal H}^{(3)}= \frac{\as}{2\pi}\left\{(1+1/N_c)\left[H_{qq}(J_{12})+
H_{qq}(J_{23})+ H_{qq}(J_{31})\right]+3C_F/2\right\}
\label{H-qqq}
\ee
with $H_{qq}$ given by \re{H-qq}. 
The Schr\"odinger equation \re{gamma} with the
Hamiltonian defined in this way has a hidden integral of motion
\be
q=i\left(\partial_{x_1}-\partial_{x_2}\right)
\left(\partial_{x_2}-\partial_{x_3}\right)
\left(\partial_{x_3}-\partial_{x_1}\right) x_1 x_2 x_3
\label{Q}
\ee
and, therefore, it is completely integrable. This operator is dual to the
operator $q_3$ for the $N=3-$reggeon states, Eq.~\re{q}, in the same fashion as
the $SL(2)$ generators were related in the two cases. Similar to the Regge case,
one can identify \re{H-qqq} as the Hamiltonian of a quantum XXX Heisenberg magnet
of $SL(2,\mathbb{R})$ spin $j_q=1$. 
The number of sites is equal to the number of quarks.

Based on this identification we shall argue now that the calculation of the
anomalous dimensions can be formulated entirely in terms of Riemann surfaces
which in turn leads to a stringy/brane picture. It is important to stress here
the key difference between Regge and light-cone limits of QCD. 
In the first case the impact parameter space provides the complex plane for 
the Reggeon coordinates and we are dealing with a $(2+1)-$dimensional 
dynamical system. In the second case the QCD evolution occurs along 
the light-cone direction and is described by
a $(1+1)-$dimensional dynamical system. As a consequence, in these two cases we
have two different integrable magnets: the $SL(2,\mathbb{C})$ magnet for the
Regge limit and the $SL(2,\mathbb{R})$ magnet for the light-cone limit. The
evolution parameters (``time'' in the dynamical models) are also different: the
rapidity $\ln s$ for the Regge case and the RG scale $\ln\mu$ for the anomalous
dimensions of the conformal operators.

Our approach to calculation of the anomalous dimensions via Riemann surfaces
looks as follows. For concreteness, we shall concentrate on the evolution kernel
\re{H-qqq}. Similarly to the Regge case, one starts with the finite-gap solution
to the classical equation of motion of the underlying $SL(2,\mathbb{R})$ spin
chain and identifies the corresponding Riemann surface
\be
\omega - \frac{x^6}{\omega} = 2 x^3 -(N+2)(N+3) x + q\,,\qquad
\omega=x^3 \e^p
\ee
where $q$ is the above mentioned integral of motion \re{Q} and $N$ is the total
$SL(2,\mathbb{R})$ spin of the magnet, 
or equivalently the number of derivatives
entering the definition of the conformal operator \re{conf-op}. In distinction
with the $SL(2,\mathbb{C})$ case, Eq.~\re{h}, the total spin $N$ takes
nonnegative integer values. Note that the  Riemann surface corresponding 
 to  the three-quark operator  has genus $g=1$, while $g=0$
 for the twist 2 operators.

At the next step we quantize the Riemann surface as follows. We replace
$p=i\partial/\partial x$ and impose the equation of the complex curve as the
operator annihilating the Baxter function
\be
\lr{\e^{i\partial/\partial x}+\e^{-i\partial/\partial x}}x^3 Q(x)=
\left[2 x^3 -(N+2)(N+3) x + q\right]Q(x)\,.
\label{curve-R}
\ee
This leads to the Baxter equation for the 
Heisenberg $SL(2,\mathbb{R})$ magnet of spin $j=1$
\beq
(x+i)^3 Q(x+i)+(x-i)^3Q(x-i)=
\left[2 x^3 -(N+2)(N+3) x + q\right]Q(x)\,.
\label{bax}
\eeq
Similar to the Baxter equation in the Regge case, this equation does not have a
unique solution. To avoid this ambiguity one has to impose the additional
conditions that $Q(x)$ should be polynomial in $x$. This requirement leads to the
quantization of the integral of motion. The resulting polynomial solution
$Q=Q_q(x)$ has the meaning of the one-particle wave function in the separated
variables $x$ which in the case of the $SL(2,\mathbb{R})$ magnet take arbitrary
real values.

Given the polynomial solution to the Baxter equation \re{bax}, one can determine
the eigenspectrum of the Hamiltonian \re{H-qqq} and in result the anomalous
dimensions of the corresponding baryon operators
\beq
\gamma^{(3)}_{N,q} =\frac{\as}{2\pi}\left[(1+1/N_c){\cal E}_{N,q}+3C_F/2\right]\,,\qquad
{\cal E}_{N,q}=
i\frac{Q_q'(i)}{Q_q(i)} - i\frac{Q_q'(-i)}{Q_q(-i)}
\label{energy}
\eeq
parameterized by the eigenvalues of the integral of motion \re{Q} given by
\be
q = 
-i\frac{{Q_q}(i)-{Q_q}(-i)}{Q_q(0)}
\ee

The solution of the Baxter equation simplifies greatly in the quasiclassical
approximation which is controlled by the total $SL(2,\mathbb{R})$ spin of the
system $N$. For $N\gg 1$ the spectrum of the integral of motion $q$ is determined
by the WKB quantization condition~\cite{kor,K-dual}
\be
\oint_A p\,dx =2\pi (n+1/2)+{\cal O}(1/N)
\label{WKB-R}
\ee
where $p$ was introduced in \re{curve-R}. Here integration goes over the
$A$-cycle on the Riemann surface defined by the spectral curve \re{curve-R}. This
cycle encircles the interval on the real $x-$axis on which $|\e^p|>1$. Solving
\re{WKB-R} one gets
\be
q=\pm \frac{N^3}{\sqrt{27}}
\left[1-3\left(n+\frac12\right)N^{-1} +\CO(N^{-2})\right]\,.
\label{sol-R}
\ee
Comparing this expression with \re{AD-point} and taking into account that in this
case $q_2=-(N+3)(N+2)$ and $q_3=q$ we conclude that for $N\to\infty$ the system
is approaching the Argyres-Douglas point. Note also that the WKB quantization
conditions \re{WKB-R} involve only the $A-$cycle on the Riemann surface and
unlike the Regge case there is no $S-$duality in the quantum spectrum.

Finally, the spectrum of the anomalous dimensions in the WKB approximation  is
given by~\cite{kor,K-dual,bm}
\be
{\cal E}_{N,q} = 2\ln 2 - 6 + 6\gamma_E + 2 {\rm Re} \sum_{k=1}^3
\psi(1+i\delta_k) + \CO(N^{-6}),
\label{Energy}
\ee
where $\delta_k$ are defined as roots of the following cubic equation:
\be
2\delta_k^3 - (N+2)(N+3)\delta_k + q=0
\ee
and $q$ satisfies \re{sol-R}.

What can we learn about  stringy picture from this information about anomalous
dimensions? Let us remind that in the spirit  of  string/gauge fields
correspondence \cite{polyakov} the anomalous dimensions of gauge field theory
operators coincide with excitation energies of a string in some particular
background. An important lesson that we have learned from the analysis of the
the two- and three-quark operators is that in the first case the 
anomalous dimensions are uniquely specified by a {\it single\/} parameter 
$N$ which define the total $SL(2,\mathbb{R})$ spin. 
In the second case, a new quantum number emerges due to
the fact that the corresponding dynamical system is completely integrable. The
additional symmetry can be attributed to the operator $q$ defined in \re{Q}.
{}From the point of view of a classical dynamics this operator generates the
winding of a particle around the $A-$cycles on the spectral curve. Within the
string/gauge fields correspondence one expects to reproduce these properties
using a description in terms of the same string propagating in different
backgrounds. One is tempting to suggest that different properties of the
anomalous dimensions of the two- and three-particle opereators 
 should be attributed to different properties of the background. 
In the case of the twist-2 the anomalous dimensions
depend on integer $N$ which in the classical system has an interpretation 
of the total $SL(2,\mathbb{R})$ angular momentum of the system. 
On the stringy side the same parameter has a natural interpretation as a 
string angular momentum.

When this paper was in a process of preparation the interesting paper concerning
the stringy derivation of the anomalous dimensions of the twist-2 operators
\cite{gkp} appeared.%
\footnote{Note that some attempts to develop the $\sigma$-model representation
for high energy QCD were undertaken a long time ago \cite{verlinde}.} It was
shown that the classical solution to the equation of motion in the $\sigma$-model
on $AdS_5
\times S^5$ provides the anomalous dimension of the twist-2 operators with the
large spin $N$. Since our derivation of the  same anomalous dimension is
seemingly different it is very instructive to compare two approaches. The logic
behind derivation in \cite{gkp} implies that one considers the stringy $\sigma$
model perturbed by some vertex operator representing the operator under
consideration on the gauge theory side. Then the energy on the solution to the
classical equations of motion in $\sigma-$model involving the radial coordinate
in the bulk provides the anomalous dimension of the operator.

In our approach we have the following correspondence
\ba
\mbox{operator} &\Longleftrightarrow& \mbox{Riemann surface}
\nonumber
\\
\mbox{twist of the operator} &\Longleftrightarrow& \mbox{genus of
the Riemann surface}
\nonumber
\\
\mbox{calculation of the anomalous dimension}
&\Longleftrightarrow& \mbox{quantization of the Riemann surface}
\nonumber
\ea
It seems that the Riemann surfaces whose moduli (the integrals of motion of the
spin chain) define the anomalous dimensions of the corresponding operators
describe the $\sigma-$model solutions found in \cite{gkp}. The precise relation
between two approaches need to be clarified further and will be discussed
elsewhere.

As we have seen the quantization of the Riemann surface can be performed most
effectively in terms of the Baxter equation. It is worth noting that the solution
to the Baxter equation can be identified as a wave function of D0 brane
\cite{gnr}. Quantization conditions arise from the requirement for the wave
function the D0 brane probe in the background of the Riemann surface to be well
defined.

In the case of the light-cone composite operators we have to incorporate into the
stringy picture a new quantum number which is parameterized by an integer $n$,
Eq.~\re{sol-R}, i.e. string excitation spectrum has now different sectors
parameterized by this integer. The natural way to interpret these sectors is to
identify $n$ with the winding number of a closed string. The corresponding
background for such scenario is offered by the Riemann surface itself with the
string wrapped around the $A-$cycle. It is an interesting open question if one
can interpret a momentum in WKB quantization condition \re{WKB-R} as a momentum
of a string  $T$-dual to the string with the winding number $n$.

Since the spectrum of the anomalous dimensions in QCD coincides with the spectrum
of the $SL(2,\mathbb{R})$ spin chain Hamiltonian it would be interesting to
explore further the symbolic relation
\beq
H_{\rm string}\propto H_{\rm spin}
\eeq
where string propagates in the background determined by the Riemann surface of
the spin chain. It is known that hamiltonian formulation of the spin chains is
closely related  to the Chern-Simons (CS) theory with the inserted Wilson lines.
The number of the sites in the spin chain $N$ corresponds to the number of the
Wilson lines. The gauge group in the CS theory is the symmetry group of the
magnet. Here it is the $SL(2,\mathbb{R})$ and group manifold is an $AdS_3$ space.
Hence the spin chain describes the motion of $N$ points in $AdS_3$ space.

\section{On the dual bulk representation of the reggeon}

Let us make some comments about a dual gravity description of multi-colour QCD in
the Regge limit. We can not present the complete solution of the problem due to
lack of the proper (super)gravity background for nonsupersymmetric theories but
mention some features which seem to be important for the behaviour of the
scattering amplitudes. First of all we want to emphasize that in spite of the
large $N_c$ limit one has to deal with Riemann surfaces whose genus is determined
by the number of reggeons. In the scattering process the relevant geometry is
captured by the M2 brane wrapped around the Riemann surface in the background
Taub-NUT metric. Two transverse coordinates and two momenta are involved in this
Taub-NUT background. It is this part of the total metric which provides the
intercepts of the multi-reggeon states. We expect that back reaction of
``scattering''M2 brane on the ``vacuum'' M5 brane can be neglected.

The M2 brane modifies the background and it is known that its near horizon
geometry involves $AdS_4$ factor. One of the M2 coordinates is the time
direction. Because here we  are dealing with static  properties of the system one
can study them at any moment in time. As a consequence we are interested in the
fixed time near horizon geometry which gives  us $AdS_3$ space. This  could help
 to  determine  a proper CFT  on  the transverse plane \cite{boot} but   an exact
 respective  conformal field theory is still unknown and  CFT interpretation of
reggeized gluons  and their compound states in pomeron and odderon sectors
remains  obscure.

Let us formulate here   a conjecture  that in  an AdS/CFT description 
reggeized gluons are  described by  $AdS_3$ singletons \cite{singleton}. 
This conjecture is based on the observation that  the correlation function  of two reggeized gluons are logarithmic in transverse space
 and they can be identified with zero dimension logarithmic operators.
 It was shown some time ago  that in  the AdS/CFT correspondence the
logarithmic operators on the boundary correspond  to the singletons in
the bulk \cite{Kogan:1999bn}.

To see the logarithmic correlator to appear let us consider the amplitude 
for the scattering of two quarks with infinite energy
\be
W^{i'j'}_{ij} \equiv \bra{0} {\cal T} W_+^{i'i}(0) W_-^{j'j}(z)\ket{0}\,.
\lab{W1}
\ee
Here, Wilson lines $W_+$ and $W_-$ are evaluated along infinite lines in the
direction of the quark velocities $v_1$ and $v_2$, respectively:
\ba
W_+(0)= P\exp \left(
       i\int_{-\infty}^{\infty} d\alpha\  v_1\cdot A(v_1\alpha)\right)\, ,
\qquad \nonumber \\
W_-(z)= P\exp\left(
       i\int_{-\infty}^{\infty}d\beta\ v_2\cdot A(v_2\beta+z)\right)\, ,
\lab{W-def}
\ea
and the integration paths are separated by the impact vector $z=(0^+,0^-,\vec z)$
in the transverse direction.

To understand the $z-$dependence, let us consider as an example
the one-loop calculation of $W$ in the Feynman gauge. One gets
\ba
W_{1-loop}=I\otimes I+(t^a\otimes t^a)\frac{g^2}{4\pi^{D/2}}\Gamma(D/2-1)
\lambda^{4-D} \nonumber \\
\int_{-\infty}^{\infty}d\alpha\int_{-\infty}^{\infty}d\beta
\frac{(v_1v_2)}{[-(v_1\alpha-v_2\beta)^2+{\vec z\,}^2+i0]^{D/2-1}}
\lab{W-one}
\ea
In this expression,
gluon is attached to both Wilson lines at points $v_1\alpha$ and $v_2\beta+z$ and
we integrate the gluon propagator $v_1^\mu v_2^\nu
D_{\mu\nu}(v_1\alpha-v_2\beta-z)$ over positions of these points. To regularize
IR divergences we introduced the dimensional regularization with
$D=4+2\varepsilon,$ ($\varepsilon > 0$) and $\lambda$ being the IR
renormalization parameter. There is another contribution to $W_{1-loop}$
corresponding to the case when gluon is attached by both ends to one of the
Wilson lines. A careful treatment shows that this contribution vanishes.
Performing integration over parameters $\alpha$ and $\beta$ we get
\be
W_{1-loop}(\gamma,\lambda^2{\vec z\,}^2) =I\otimes I+(t^a\otimes t^a)
\frac{\as}{\pi}
(-i\pi\coth\gamma)
\Gamma(\varepsilon)
(\pi\lambda^2{\vec z\,}^2)^{-\varepsilon}
\lab{W-z}
\ee
where $(v_1 v_2) = \cosh\gamma$.

The integral over $\alpha$ and $\beta$ in \re{W-one} has an infrared divergence
coming from large $\alpha$ and $\beta$. In the dimensional regularization, this
divergence appears in $W_{1-loop}$ as a pole in $(D-4)$ with the
renormalization parameter $\lambda$ having a sense of an IR cutoff.
We see that in an octet channel we got a logariphmic correlation which
 depends only on two transverse coordinates
\ba
(t^a\otimes t^a)\frac{\as}{\pi} (i\pi\coth\gamma)\ln (\pi\lambda^2{\vec z\,}^2)
\ea
which allows us to make a conjecture that  reggeized gluon is described by some
logarithmic operator with zero dimension which correspond to a singleton field in
$AdS_3$.

The arguments above suggest that the CFT at the transverse plane could be
logarithmic \cite{LCFT}. Also it is tempting to relate the $SL(2,\mathbb{C})$
symmetry of the $AdS_3$ space with the $SL(2,\mathbb{C})$ symmetry of  the
original XXX spin chain but at this stage we do not have reasonable arguments in
favor of this connection. Certainly these important issues deserve further
investigation.

\section{Conclusions}

Let us  summarize the obtained results. In this paper we proposed the stringy
picture for multi-colour QCD in the Regge limit and on the light-cone in which
hidden integrability plays the central role. 
Our approach is similar in many
respects to the one which is known  to be very successful in 
 the description of the low energy limit of the $\cal{N}$=2 SUSY YM theories. 
In the Regge limit we encounter  a new situation in comparison with a SUSY case.
One has  to develop the
quantum picture involving the dynamics on a Riemann surface as the quasiclassical
limit. We argued that the whole configuration relevant in the Regge limit follows
from  M2 brane wrapped around the spectral curve of the integrable system. Within
our approach  the Regge limit   turns out to be in the same universality class as
$\cal{N}$=2 superconformal SUSY YM  with $N_f=2N_c$ at the strong coupling
regime. In both cases there is no natural place for the $\Lambda_{QCD}$ type
parameter: in SUSY case theory is conformal while in the Regge case we are
dealing only with perturbative regime in  generalized leading logarithmic
approximation.

The WKB quantization conditions  providing the compound multi-reggion spectrum
and the spectrum of anomalous dimensions for composite light-cone operators  are
formulated in the stringy terms. They  imply  that energy of  strings properly
wrapped around the spectral curve has to be quantized. Note that since the
quantization of the integrable system amounts to the quantization of the moduli
space of the complex structures of the Riemann surface quantum gravity in two
dimensions is involved. The correspondence $\mbox{operator}
\Leftrightarrow \mbox{Riemann surface}$  plays an  important role
in our approach and is some new step toward a complete formulation of the
gauge/string correspondence.

One more issue which certainly deserves further investigation is the meaning of
the unitary Froissart boundary. Let us emphasize that  to get the
 scattering amplitude one has to sum over all  Riemann surfaces of arbitrary genus,
corresponding to the different number of Reggeons $N$
\be
A(s,t)= \sum_{N\ge 2} A_{N}(s,t) = \sum_{\rm genus=N-2}A_{g}(s,t)
\ee
in perfect agreement with the string picture. Hence, if the picture suggested for
the Froissart saturation in \cite{giddings} is correct it would mean that the
process of the black hole creation in the $s$-channel is equivalent to the closed
string propagation in nontrivial background in the $t$-channel.

Let us mention a few other important questions which remain  open. One of  them 
 is to  find  more clear physical identification of the  M-theory 10-th dimension 
involving the geometry of the spectral curve. At a moment we can accurately introduce 
it only via the Bloch-Floquet factor of the
Baker-Akhiezer function  but more physical interpretation is highly desirable.
The related question concerns the interpretation of the ``magnetic'' quantum
numbers amounted from the spectral curves. It would help in a wider application
of the dualities in the Regge limit  and possibly  could shed the additional
light on the $s-t$ duality in the high energy amplitudes.

We are grateful to S.~Derkachov, Yu.~Dokshitzer, A.~Kaidalov, A.~Leonidov,
A.~Manashov, N.~Nekrasov and F.~Smirnov for the useful discussions. The work of
A.G. is partially supported by grants CRDF-01-2108 and INTAS-00-334,   I.K. is
supported in part  by PPARC rolling grant PPA/G/O/1998/00567  and EC TMR grant
HPRN-CT-1999-00161, both I.K. and G.K.  are partially supported  by a joint
CNRS-Royal Society grant.

\end{document}